\documentclass{mn2e}
\usepackage{epsfig}
\usepackage{amsmath}
\usepackage{textcomp}
\usepackage{longtable}
 
\def\d{\mbox{d}}

\setlongtables

\title[Modelling the X-ray spectra of high velocity outflows from quasars]{Modelling the X-ray spectra of high velocity outflows from quasars}
\author[S. A. Sim]{S. A. Sim\thanks{s.sim@imperial.ac.uk} \\
Astrophysics Group, Imperial College London,
Blackett Laboratory, Prince Consort Road, London, SW7 2AZ, UK}

\date{\today}

\pagerange{\pageref{firstpage}--\pageref{lastpage}}
\pubyear{2004}

\begin{document}
\maketitle
\label{firstpage}

\begin{abstract}
High velocity outflows from supermassive
black holes have been invoked to explain the recent identification of strong
absorption features in the hard X-ray spectra of several quasars.
Here, Monte Carlo radiative transfer calculations are performed to synthesise
X-ray spectra from models of such flows.
It is found that simple, parametric bi-conical outflow models with plausible choices for the
wind parameters predict spectra that are in good qualitative
agreement with observations in the 2 -- 10~keV band.
The influence on the spectrum of both the mass-loss rate and opening angle of the flow are
considered: the latter is important since photon leakage plays a significant role
in establishing an ionization gradient within the flow, a useful discriminant between 
spherical and conical outflow for this and other applications.
Particular attention is given to the bright quasar PG1211+143 for which
constraints on the outflow geometry and mass-loss rate 
are discussed subject to the limitations of the currently available
observational data.
\end{abstract}


\section{Introduction}

There is a growing body of evidence for high-velocity, blueshifted 
absorption features in the hard (2 -- 10~keV) X-ray spectra of quasars.
Chartas et al. (2002) first identified a pair of absorption features in the
X-ray spectrum of the relatively high redshift quasar APM08279+5255 
($z=3.91$)
and proposed that they were due to highly ionized iron. 
They were unable, however, to
identify the ionization stage with certainty.
More recently, Pounds et al. (2003a) have identified narrow absorption features
in the spectrum of the nearby quasar PG1211+143. In this case, the 
identification
of an X-ray line at $\sim 7.5$~keV with the
significantly 
blueshifted ($v \sim 0.1c$) 
K~$\alpha$ transition of
Fe~{\sc xxv} or {\sc xxvi} was supported by the identification of other, 
weaker features at softer energies.
Similar narrow absorption features have also now been reported in a second 
high-redshift quasar (PG1115+080, Chartas, Brandt \& Gallagher 2003) and a
second low-redshift quasar (PG0844+349, Pounds et al. 2003b). In addition, 
high velocity
absorption has been reported in yet another quasar (PDS456) by
Reeves, O'Brien \& Ward (2003), although in this case the absorption is
broader and the authors favour an interpretation involving absorption edges
of Fe~{\sc xvii}-{\sc xxiv} rather than lines of more highly ionized material.

It has been suggested that 
these absorption features form in a fast, massive 
outflow from the accreting supermassive black hole
in the quasar nucleus (Chartas et al. 2002, 
Pounds et al. 2003a, King \& Pounds 2003).
To account for the observed absorption features requires such flows to be
both more highly ionized and have significantly greater column densities than
necessary for 
the outflows detected via blueshifted ultraviolet absorption lines
in broad absorption line (BAL) quasars (Pounds et al. 2003b). 
It is not yet established whether 
highly ionized fast outflows might be common in the quasar population, 
but it is clear that
when present, they may carry sufficient energy to be important in the 
energy budget of  
accretion by the nuclear black hole.
In addition,
King \& Pounds (2003) have presented simple arguments to show that
outflows from black holes accreting at around the Eddington limit are likely
to be optically thick giving rise to an XUV photosphere, emission from which
may be responsible for the ``big blue bump'' (BBB) seen in quasar spectra.

This interpretation, however, is not firmly established: in some instances (e.g. 
APM08279+5255; Chartas et al. 2002) there remains ambiguity in the line identification
while in others (PG1211+143, PDS456; McKernan, Yaqoob \& 
Reynolds 2004) it has been suggested that 
some portion of 
the absorption features is due to
gas in the vicinity of our Galaxy, rather than material 
intrinsic to the quasar. In addition, Kaspi (2004) has shown that many of the
features in the spectrum of PG1211+143 can be explained by an alternative model 
involving a relatively small outflow velocity ($\sim 3000$~km~s$^{-1}$). It is noted,
however, that the preliminary study presented by Kaspi (2004) addresses only the
{\it XMM-Newton}
Reflection Grating Spectrometer (RGS) data and not the European Imaging Camera (EPIC) data:
it is not clear that a low-velocity outflow could explain the absorption features in the EPIC
spectrum identified by Pounds et al. (2003a).

To better understand the hard X-ray absorption features and the potentially
important outflows with which they may be associated requires 
synthesis of the X-ray spectrum predicted by 
plausible physical models.
The primary goal of this paper is to perform
realistic radiative transfer calculations and synthesise spectra in order to
determine whether simple 
physical 
outflow models, as discussed by Pounds et al. (2003a) and King \&
Pounds (2003), readily predict strong, narrow absorption features as required
by the observations.

A secondary objective is to use model spectra to constrain the plausible
range of flow parameters by comparison with the data for a particular quasar.
This investigation is focused on the narrow emission line quasar
PG1211+143. Of the two nearby quasars 
in which narrow hard X-ray absorption lines
have been detected by Pounds et al. (2003a,b), this object has been chosen
over the other candidate (PG0844+349)
since the observational data is of higher quality and 
a wider variety of absorption features have been identified. 
PG1211+143 is bright, nearby ($z = 0.0809$, Marziani et al.
1996) and known to be radiating at around the Eddington luminosity
(Boroson 2002, Gierli\'{n}ski \& Done 2004). 
Based on simple ionized absorber fits to the data, Pounds et al. (2003a) 
suggest the high-velocity outflow 
in PG1211+143 has a mass-loss rate 
$\sim 3$~M$_{\odot}$~yr$^{-1}$, and they suggest that the outflow is
likely to subtend a wide opening angle as viewed from the central
black hole.

This work is also of general interest for the study of radiative transfer in non-spherical outflows. 
It will be shown that calculations for conical geometries are significantly 
different from spherical outflow, primarily owing to the influence of photon leakage through
the conical boundary.

During the refereeing of this article, Everett \& Ballantyne (2004)
presented a related study on radiatively driven outflows from black holes.
They investigated continuum driven 
flows and conclude that, while such flows are possible, they are likely to be too
highly ionized to produce absorption line features. 
They did not, however, investigate line or MHD driven outflows nor consider 
multi-dimensional radiative transfer effects.

In Section 2, the outflow model that will be considered is
discussed.
The calculations presented in this paper have been performed with a Monte 
Carlo 
radiative transfer code which has been adapted from the code written by
Sim (2004). The code uses the Macro Atom formalism developed by Lucy (2002,
2003) and is discussed in Section 3. 
Section 4 identifies the atomic data used in the Monte Carlo simulations.
The spectra computed from various models are discussed in 
Section 5 and conclusions drawn in Section 6.

\section{Model}

The geometry adopted in this investigation is that of a bi-conical
outflow as discussed by Pounds et al. (2003a) and King \& Pounds (2003).
The flow has opening angle $\theta_0$ and
is launched at radius $R_{c}$ from the supermassive black hole at 
the core of the quasar. Figure~1 shows a cartoon of this flow geometry.
Note that in the limit $\theta_0 = 90$\textdegree~a spherical wind with
no obscuring material is recovered.
In the subsections below, the various physical quantities which 
are required for the radiative transfer calculations are discussed.

\begin{figure}
\hspace*{1cm}
\epsfig{file=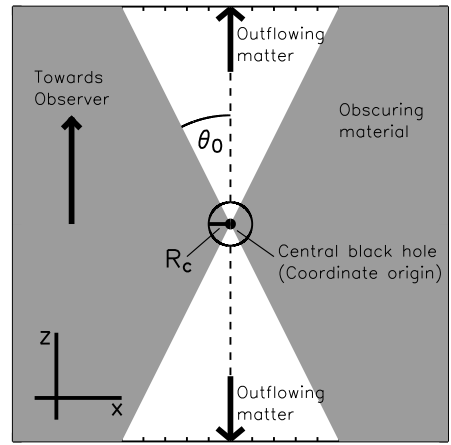, width=9cm}\\
\caption{The bi-conical flow geometry under consideration. The outflow 
(unshaded) has
opening angle $\theta_0$ and is launched at radius $R_{c}$ from the 
central black hole.
The shaded region is assumed to contain obscuring material through which photons
do not propagate. 
Note that in the limit $\theta_{0} = 90$\textdegree~a spherical outflow
with no obscuring material is obtained.
The observer is assumed to be at infinity along the positive
$z$-axis. The flow is symmetric under rotation about the $z$-axis.}
\end{figure}

\subsection{Velocity and Density}

It is assumed that the velocity of the flow increases monotonically outward 
and is
a function of radius ($r$) alone. A simple $\beta$-type velocity law (of the
form commonly used in modelling stellar winds) is adopted

\begin{equation}
v (r)  =  v_{c} + (v_{\infty} - v_{c}) \left( { 1 - \frac{R_{c}}{r}} 
\right)^{\beta}
\end{equation}
where $v_c$ is the outflow velocity at the base of the wind (i.e. at $r = R_{c}$) and
is assumed to be negligibly small;
$v_{\infty}$ is the 
terminal velocity of the flow, which is constrained by the 
observed line shifts; and $\beta$ controls the rate of
acceleration in the wind.

Given the velocity, the mass density $\rho(r)$ is determined by the
equation of mass conservation for an outflow with mass-loss rate $\Phi$

\begin{equation}
\rho(r) = \frac{\Phi}{4 \pi v(r) r^2 b}
\end{equation}
where $b = 1 - \cos \theta_0$ 
is the fraction of solid angle subtended
by the flow as viewed from the coordinate origin.

\subsection{Ionization}

The ionization state of all the elements included in the calculation is
computed from the radiation field on the assumption of
ionization equilibrium. During the Monte Carlo simulations,
estimators for the photoionization rate in each bound-free continuum
are obtained (following Lucy 2003, see Section 3) 
and used to compute the ionization state
by balancing against radiative recombination rates. Collisional ionization is
neglected -- this is expected to be a good approximation in view of the 
intense radiation field and the
high ionization potential of the ions important to this study.

The radiative recombination rates depend on the local electron energy
distribution. Since thermal balance is not pursued here, this energy
distribution is not calculated in detail but is assumed to be Maxwellian
with a fixed electron temperature, $T_{e}$. The adopted value of $T_{e}$ is
discussed in Section 5.

\subsection{Excitation}

All the X-ray spectral lines of interest are ground state transitions and
so the excitation of levels above the ground state has little effect
on the calculations. Therefore the following prescription for excitation is
adopted

\begin{equation}
\begin{array}{cccc}
{n_{i}}/{n_{g}} &=& \left( {{n_{i}}/{n_{g}} }\right)^{*}_{T_{r}} & 
\mbox{if level $i$ is metastable} \\
\\
{n_{i}}/{n_{g}} &=& W(r) \left( {{n_{i}}/{n_{g}}} \right) ^{*}_{T_{r}} & 
\mbox{otherwise}
\end{array}
\end{equation}
where $n_{i}$ is the level population (number density) for an excited
state $i$, $n_{g}$ is the population of the ground state and $T_{r}$ is
a chosen radiation temperature.
The geometric dilution factor is given by
$W(r) = \frac{1}{2}(1 - \sqrt{ 1 - (R_{c}/r)^2})$ 
and the notation $( \; )^*_{T_{r}}$ indicates that 
the quantity in parenthesis
should be evaluated in LTE at temperature $T_{r}$. 
In practice, 
$T_{r}$ is set equal to $T_{e}$ throughout.

\subsection{Incoming broadband spectrum}

\begin{figure}
\epsfig{file=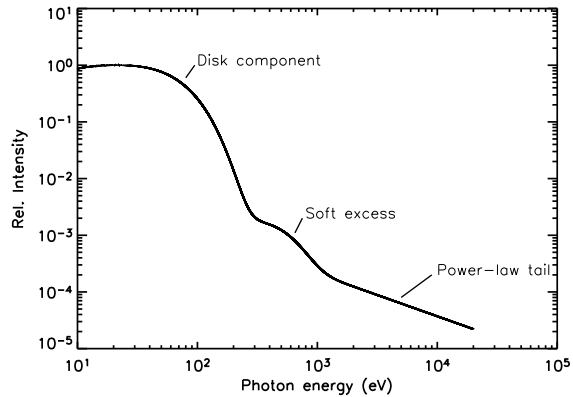, width=8cm}\\
\caption{The form of the input spectrum of radiation energy. The three distinct components
are indicated in the figure.}
\end{figure}

The Monte Carlo simulations require that the incoming energy spectrum at
the base of the wind be specified. In this paper, the incoming
spectrum consists of three distinct components (see Figure 2). Energetically, the dominant
component is that of the accretion disk around the central black hole.
This component is modelled by a simple multi-colour black-body 
spectrum (Mitsuda et al. 1984). The spectral energy distribution of the 
disk was obtained from equation~4 of Mitsuda et al. (1984) with the
disk inner radius $r_{\mbox{\scriptsize in}} = 3 R_{s}$ and the outer
radius $r_{\mbox{\scriptsize out}} = 100 R_{s}$ ($R_{s}$ is the
Schwarzchild radius of the black hole). Since PG1211+143 is believed to be
accreting at, or close to, the Eddington limit (Boroson 2002, Gierli\'{n}ski 
\& Done 2004), 
the total disk luminosity was fixed at the Eddington luminosity.
The mass of the central black hole, which is required for the
calculation of $R_{s}$ and the Eddington luminosity, 
is taken to be $4 \times 10^{7}$M$_{\odot}$
(Kaspi et al. 2000).

At hard X-ray wavelengths the continuum spectrum is well described by a 
simple power law which most probably arises through inverse
Compton scattering of seed disk photons in an optically thin region 
in or near the disk. The formation of this power law is not addressed here 
but it is included as part of the input spectrum. The photon power law
index $\Gamma = 1.75$ is adopted (the observed index for PG1211+143 has
been determined by Pounds et al. [2003a] from both {\it XMM-Newton} pn and
MOS data yielding respectively $\Gamma \sim 1.79$ and $\Gamma \sim 1.71$).
The normalisation of the power-law component relative to that of the disk
is chosen in order to obtain
consistency with the observed X-ray (2 -- 10 keV) luminosity.
To achieve this, the normalisation is adjusted iteratively in the Monte Carlo
calculations (see Section 3.6).

At soft X-ray energies a significant excess of emission is observed relative to the
power-law component (Pounds et al. 2003a). This soft-excess
is included in the input spectrum here. Its shape is modelled by a black-body 
with temperature $T = 1.3 \times 10^6$~K, the temperature of the primary 
black-body component fitted to the spectrum by Pounds et al. (2003a). The 
normalisation of this component is fixed to that of the power-law by the
observation that the intensity of the soft-excess component is approximately
equal to that of the hard power-law at $\sim 0.8$~keV (see figure 3 of Pounds et
al. 2003a). An excess of emission 
in the soft X-ray band
is common in quasars but its origin is not
well understood (see e.g. Gierli\'{n}ski \& Done [2004] for a recent discussion). For the purposes of this
paper, however, the soft excess is of relatively little concern since it is not
a major contributor to the hard X-ray region.

\section{Monte Carlo simulations}

The code used to perform radiative transfer calculations is based on that
discussed by Sim (2004) which uses the Macro Atom formalism recently
developed by Lucy (2002,2003). The principles of the code are briefly 
discussed below, with particular emphasis on those aspects that 
differ from the Sim (2004) code.

\subsection{Geometry}

The code adopts a bi-conical geometry with opening angle 
$\theta_0$ (see Section 2 and Figure 1).
In practise the radiative transfer 
calculation is performed in one cone and then the relevant parameters of the flow
(e.g. the mass-loss rate $\Phi$) are doubled to account for the bi-conical
structure.
Energy packets are launched through the surface at the base of the cone  
(radius $R_c$ from the black hole) and their
propagation through the cone is tracked in both radial ($r$) and angular 
($\theta$) position until they leave the flow through the outer boundary at 
$r = r_{\mbox{\scriptsize max}}$ or the conical boundary at $\theta = \theta_{0}$.
The lower boundary $r=R_{c}$ is assumed to be reflecting so that no packets are
lost through this surface. Packets which pass through the outer boundary
$r = r_{\mbox{\scriptsize max}}$ are assumed to escape to infinity while those that
pass through the conical boundary are assumed to be destroyed by obscuring material
(such as might be present in the form of a dusty torus surrounding the AGN 
[Antonucci \& Miller 1985, 
Pier \& Krolik 1992], or the warm highly ionized medium [WHIM] discussed by Elvis [2000]). 
It is assumed that no photons enter the flow through the conical
boundaries. In practise there would be little effect if photons did enter via 
these boundaries provided that they are not hard enough to influence the ionization
state of the material in the flow.

\subsection{Discretization of the model}

The flow is divided into 100 cells in the radial direction equally spaced
in $x = R_{c} / r$. The outermost cell extends to $r_{\mbox{\scriptsize max}}
= 100 R_{c}$.
At present there is no angular ($\theta$) stratification.

\subsection{Propagation and interaction of packets}

The propagation of packets through the model is followed exactly as described
by Sim (2004).
The treatment of line interactions using the Sobolev approximation and the
Macro Atom formalism is also as described by Sim (2004), with the exception that
bound-free processes are now included in the Macro Atom description.
Thus, in contrast to the photoionization modelling described by Pounds et al. [2003a],
the linewidths and profiles are computed rather than pre-specified.

Following the formalism developed by Lucy (2003), the calculations presented here include
bound-free absorption and emission (including the creation and destruction of $k$-packets)
which were neglected by Sim (2004).
In addition, since X-ray photons are under investigation, electron scattering is no longer treated
in the limit of Thompson scattering but is generalised to account for Compton scattering 
as described in the next subsection.

\subsubsection{Compton scattering}

The electron scattering cross-section presented to an $r$-packet is computed using
the Klein-Nishina equation 
for the differential cross-section on the 
assumption that
the target electrons are stationary in the co-moving frame
(see e.g. Leighton 1959). 

When electron scattering occurs, the scattering angle is selected from a pre-tabulated
look-up table of the angular probability distribution as a function of incident
photon energy using the Klein-Nishina equation.

Physically, when a photon undergoes Compton scattering some fraction of its energy 
($f$, which is readily related to the photon frequency and scattering angle) is passed to 
the electron, reducing the frequency, and therefore energy, of the photon. However, such 
a splitting of energy is not desirable in the context of the ``indivisible energy packet''
approach adopted here. Therefore 
$f$ is identified as the probability that during a Compton scattering event the $r$-packet
is converted insitu to a $k$-packet (which is subsequently eliminated by thermal cooling
processes in the usual way) and $1 - f$ is the probability that the packet will continue to
propagate as an $r$-packet with diminished frequency but unchanged energy 
(in the co-moving
frame). When averaged over many interactions, this approach reproduces the physics of Compton
scattering exactly without the need to divide energy packets at any point.

It is noted that this scheme for the treatment of Compton scattering assumes that the electron 
rapidly comes back into thermal equilibrium with its surroundings following the interaction with 
the photon. This assumption is adequate here but in principle could be lifted to 
allow the energy carried by non-thermal electrons to be followed during Monte Carlo simulations
when necessary. This is likely to be important when the methods are generalised to
include inverse Compton scattering.

\subsection{Monte Carlo estimators}

For the formal calculation of the emergent spectrum (see Section 3.7), Monte
Carlo estimators are needed not only for the Sobolev mean intensity in the
spectral lines ($\bar{J}$, given by equation 10 of Sim [2004]) but also for
the mean intensity in both the far blue ($J^{b}$) and red ($J^{r}$)
wings of the lines and also for a weighted mean intensity in the far 
blue wing ($J^{b, \tau} \equiv < I^{b} e^{-\tau_{s}} >$, where $I^{b}$ is
the specific intensity in the far blue wing, $\tau_s$ is the Sobolev optical
depth and $<...>$ indicates an angular average). Estimators for these
quantities in each cell are readily constructed following Lucy (1999, 2003):

\begin{equation}
J^{b} = \frac{1}{4 \pi \Delta t} \frac{1}{V} \sum_{\mbox{\scriptsize in}} \frac{1}{\d \nu/\d s}
\epsilon
\end{equation}

\begin{equation}
J^{r} = \frac{1}{4 \pi \Delta t} \frac{1}{V} \sum_{\mbox{\scriptsize out}} \frac{1}{\d \nu/\d s}
\epsilon
\end{equation}

\begin{equation}
J^{b, \tau} = \frac{1}{4 \pi \Delta t} \frac{1}{V} \sum_{\mbox{\scriptsize in}} \frac{e^{-\tau_{s}}}{\d \nu/d s}
\epsilon
\end{equation}
In each case $\Delta t$ is the time interval represented by the Monte Carlo simulation, $V$ is the volume
of the cell in question, $\epsilon$ is the packet energy 
and $\d \nu / \d s$ is the gradient of co-moving frequency with path-length.
In the first and third cases the summation runs over all packets as they redshift {\it into} resonance 
with the line and in the second case the summation is over packets as they redshift {\it out} of resonance 
with the line.

In principle, Monte Carlo estimators for the photoionization rate and stimulated recombination rate ($\gamma$ and
$\alpha^{\mbox{\scriptsize st}}$) should be obtained using equations 44 and 45 of Lucy (2003). However, since 
the line estimators presented above provide a closely spaced grid of values for $J$ in each cell it is possible 
to obtain
$\gamma$ and $\alpha^{\mbox{\scriptsize st}}$ by integrating over this grid. This method has been adopted since 
it noticeably reduces the run time of the code (when the bound-free estimators are computed directly 
a significant fraction of computer time is spent on 
recording the large number of small contributions to these estimators which arise because of the large number of
electron scattering events). 
Although estimators obtained by this approach are slightly less accurate than those obtained with 
equations 44 and 45 of Lucy (2003), they are still of higher quality than those for the individual lines and  
since it is necessary to simulate a large enough number of packets that the line estimators are reliable, then it is
acceptable to compute the bound-free estimators from the line estimators and obtain a similar degree of accuracy. 

\subsection{The time interval}

The time interval represented by the Monte Carlo simulation, $\Delta t$ is determined by the observed
2 -- 10 keV luminosity, $L_{\mbox{\scriptsize X}} = 3.3 \times 10^{43}$~ergs~s$^{-1}$ for 
PG1211+143 (Pounds et al. 2003a).
During the Monte Carlo simulation, the total energy in the 2 -- 10 keV energy band carried by packets through the
outer boundary of the model ($E_{\mbox{\scriptsize X}}$) is recorded and used to obtain the time
interval via

\begin{equation}
\Delta t = \frac{E_{\mbox{\scriptsize X}}}{b L_{\mbox{\scriptsize X}}}
\end{equation}
where, as usual, $b$ is the fraction of solid angle subtended by the flow. The time interval ($\Delta t$)
is used both for the normalisation of the Monte Carlo estimators (see Section 3.4) and for determining the
relative normalisation of the power-law and disk components of the input spectrum in the next iteration.

\subsection{Iteration cycle}

Each complete run of the code involves several iterations of the Monte Carlo
calculation to converge the Monte Carlo estimators, ionization fractions and the
relative normalisation of the disk and power-law components of the input spectrum. Initially it is
assumed that the flow is fully ionized and a Monte Carlo simulation (with a relatively small 
number of energy packets) is performed to obtain preliminary values for the photoionization rates.
These are used to compute more realistic ionization fractions before performing additional
Monte Carlo simulations. In each successive simulation the Monte Carlo estimators 
obtained from the previous simulation are used to compute the ionization fractions and the 
Macro Atom jumping/de-activation probabilities. Also, the time interval (Section 3.5) is used to obtain the 
relative normalisation of the components of the input radiation field.
Typically five iterations were performed 
to converge the final values of the estimators to sufficient accuracy for the
spectral synthesis step described below.

\subsection{Spectral synthesis}

While it is possible to produce spectra from Monte Carlo simulations directly by examining the
frequency distribution of the energy packets emerging from the computational domain, it has been demonstrated
by Lucy (1999) that far higher quality spectra may be obtained by using the Monte Carlo estimators to 
perform a formal integral solution of the transfer equation. The accuracy of the spectrum generated in this
manner is limited not by the Monte Carlo noise in the outgoing packet frequency distribution but by the 
Monte Carlo noise in the estimators which is significantly lower.
A brief discussion of the method of spectral synthesis is given below with particular attention given to the
differences from Lucy (1999).

\subsubsection{Ray tracing}

To compute the outgoing spectrum, the specific intensity at each frequency $I_{\nu}$ is traced along a set of 
paths through the model
with a range of impact parameters.
The total emergent intensity is determined by numerical integration over
the impact parameter (see Lucy 1999). It
is assumed that the observer lies directly above the axis of symmetry of the flow (the line $\theta=0$). 
Initially,
the intensity of the ray is either set to that of the input radiation field 
(if the impact parameter is sufficiently
small that the ray trajectory originates on the inner surface boundary at $R_{c}$) or to zero 
(if the impact parameter is
large enough that the ray originates on either the conical boundary [bi-conical models] or the outer edge of the
rear hemisphere [spherical models]). The intensity of the beam is then traced along the
path to the edge of the model in the following way. First, the distance the ray must travel until the comoving
frequency of the photons with which it is associated becomes redshifted into a spectral line is found. 
The continuum 
optical depth $\tau_{c}$ and source function $S_{c}$ are computed for this path length, including contributions
from both electron scattering and bound-free processes. The continuum source function is given by

\begin{equation}
S_{c} = \frac{\tau_{es}}{\tau_{c}} S_{es}  + \frac{\tau_{bf}}{\tau_{c}} S_{bf}
\end{equation}
where $\tau_{es}$ and $\tau_{bf}$ are the optical depths due to electron scattering and bound-free processes respectively.
Since the ionization fractions
have already been computed from the radiation field and the important bound-free processes all involve recombinations from 
ground states, the bound-free source function $S_{bf}$ is computed directly from the ionization fractions in the usual 
way.
The source function for electron scattering is assumed to be isotropic  and given by the mean intensity, $J$, which is obtained by 
interpolation between the value of $J^r$ for the next spectral line to the blue and $J^b$ for the next 
spectral line to the red (in the comoving frame). The values of $J^r$ and $J^b$ are already determined by the
estimators discussed in Section 3.4.
Once the intensity is known at the point where the ray is redshifted into a spectral line, the effect of the
line on the intensity of the ray is accounted for using

\begin{equation}
I^{r}_{\nu} = I^{b}_{\nu} e^{-\tau_{s}(\mu)} + S_{bb} (1 - e^{-\tau_{s}(\mu)})
\end{equation}
where $I^{b}_{\nu}$ is the incoming intensity at the far blue wing of the line,
$I^{r}_{\nu}$ is the outgoing intensity at the far red wing of the line,
$S_{bb}$ is the
line source function and 
$\tau_{s}(\mu)$ is the
Sobolev optical depth of the line (which depends on $\mu$, the cosine of the angle between the
ray direction and the radial direction).
Since level populations are {\it not} computed self-consistently then it is not appropriate to 
compute $S_{bb}$ from the level populations but rather from the Monte Carlo estimators for the
radiation field. This is achieved by averaging the above equation over $\mu$ to obtain
an expression for $S_{bb}$ in terms of mean intensities

\begin{equation}
S_{bb} = \frac{J^{r} - J^{b, \tau}}{\frac{1}{2} \int_{-1}^{+1} ( 1 - e^{-\tau_{s}(\mu)}) \; \d\mu}
\end{equation}
Since estimators have been recorded for both $J^{r}$ and $J^{b,\tau} \equiv < I^{b}_{\nu} e^{-\tau_{s}} >$ (see section 3.4)
this allows $S_{bb}$ to be evaluated without reference to the level populations. Once the ray has passed
though resonance with the line, the distance to the next line is computed and the effects of continuum 
opacity up to that line are accounted for as before. This processes of alternately accounting for the continuum
and lines is continued until the ray reaches the top of the model whereupon the outgoing intensity is recorded and
used to compute the spectrum. In practise, since the X-ray region is relatively sparsely populated with spectral
lines, the gridding obtained for the computation of the electron scattering source function is rather coarse. This 
is easily overcome by introducing additional fake spectral lines in the Monte Carlo simulation.
These are defined to have zero optical depth and so do not affect the propagation of energy packets but they do 
cause $J$ to be computed on a finer frequency mesh, as required.

\subsection{Accuracy of the calculations}

The number of energy packets used in each simulation is chosen in order
that the Monte Carlo noise in the estimators is $< 10$~per cent. This level 
of precision is sufficient given the limited quality of the observational
data available at present. 
To this accuracy, the transformation between the
comoving and observer rest frame can be taken as Galilean and the Doppler
shift treated as non-relativistic. 
If higher accuracy were to be required it would
become necessary to consider special relativistic corrections (which are
of order $v / c \leq 0.1$). Including such corrections poses no significant
problems to the operation of the code but in view of the precision 
required here they can be neglected for simplicity.

\section{Atomic data}

Table~1 lists the elements and ions 
included in the calculations described below. For light elements only
hydrogen- and helium-like ions are included, but lower ionization stages are
included for iron and nickel.
The element abundances, relative to hydrogen, are assumed to be solar.
Atomic models, bound-bound oscillator strengths and bound-free
cross-sections have been extracted from the {\it Xstar} 
database
(Bautista \& Kallman 2001)\footnote{http://heasarc.gsfc.nasa.gov/docs/software/xstar/xstar.html}.
In the calculations, all the bound-bound transitions in the database for the 
ions listed in Table~1 are included, a total of $\sim 2500$ lines.
For hydrogen-like ions, bound-free transitions from levels with $n=1, 2, 3$ and
4 are included. For helium-like ions, bound-free transitions from the
1s$^2$, 1s2s, 1s2p and 1s3s configurations are included. For all other ions,
photoionization is included for all those of the ten lowest lying 
fine-structure energy 
levels for which cross-sections are given in the database. In total
$\sim 250$ bound-free continua are included in the calculations.

\begin{table}
\caption{The elements and ions included in the calculations.}
\begin{center}
\begin{tabular}{lccc}\hline
Element & Ions & Element & Ions\\ \hline
H & {\sc i}, {\sc ii} &  Si & {\sc xiii} -- {\sc xv}\\
He& {\sc i} -- {\sc iii} & S& {\sc xv} -- {\sc xvii}\\
C& {\sc v} -- {\sc vii} & Ar & {\sc xvii} -- {\sc xix}\\
N& {\sc vi} -- {\sc viii} & Ca & {\sc xx} -- {\sc xxii}\\
O& {\sc vii} -- {\sc ix} & Fe & {\sc xxiii} -- {\sc xxvii}\\
Ne& {\sc ix} -- {\sc xi} & Ni & {\sc xxvi} -- {\sc xxix}\\
Mg& {\sc xi} -- {\sc xiii} & \\ \hline
\end{tabular}
\end{center}
\end{table}

\begin{figure*}
\vspace*{-4.5cm}
\epsfig{file=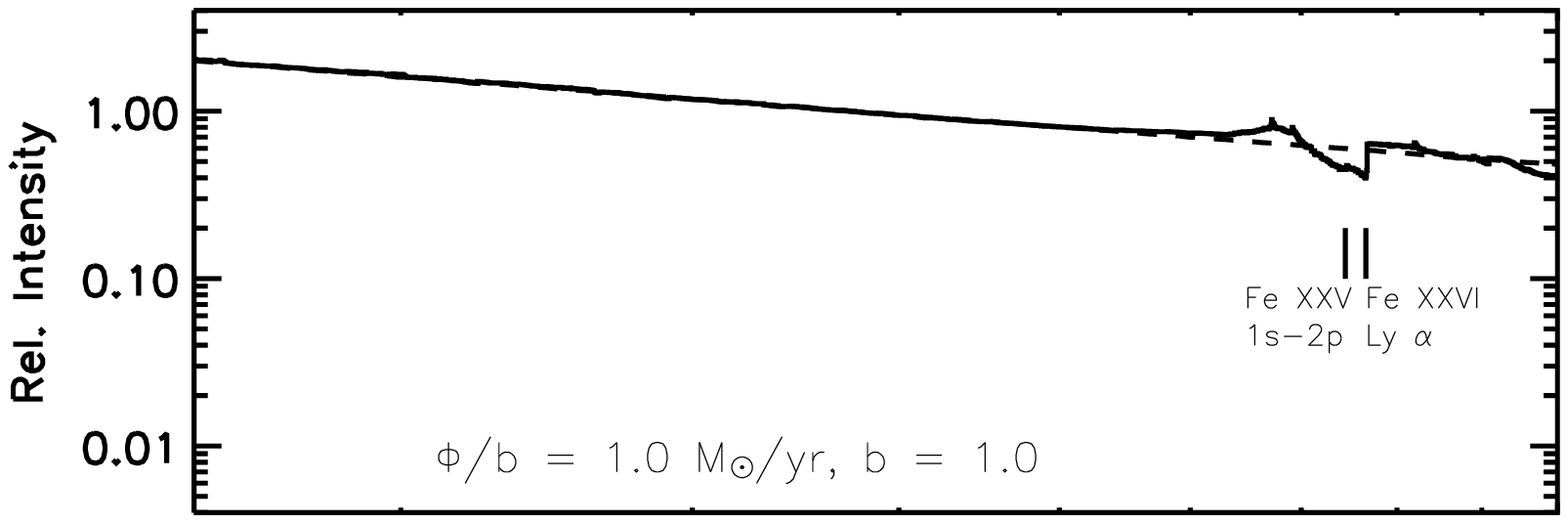, width=12cm, height=8cm}\\
\vspace*{-4.5cm}
\epsfig{file=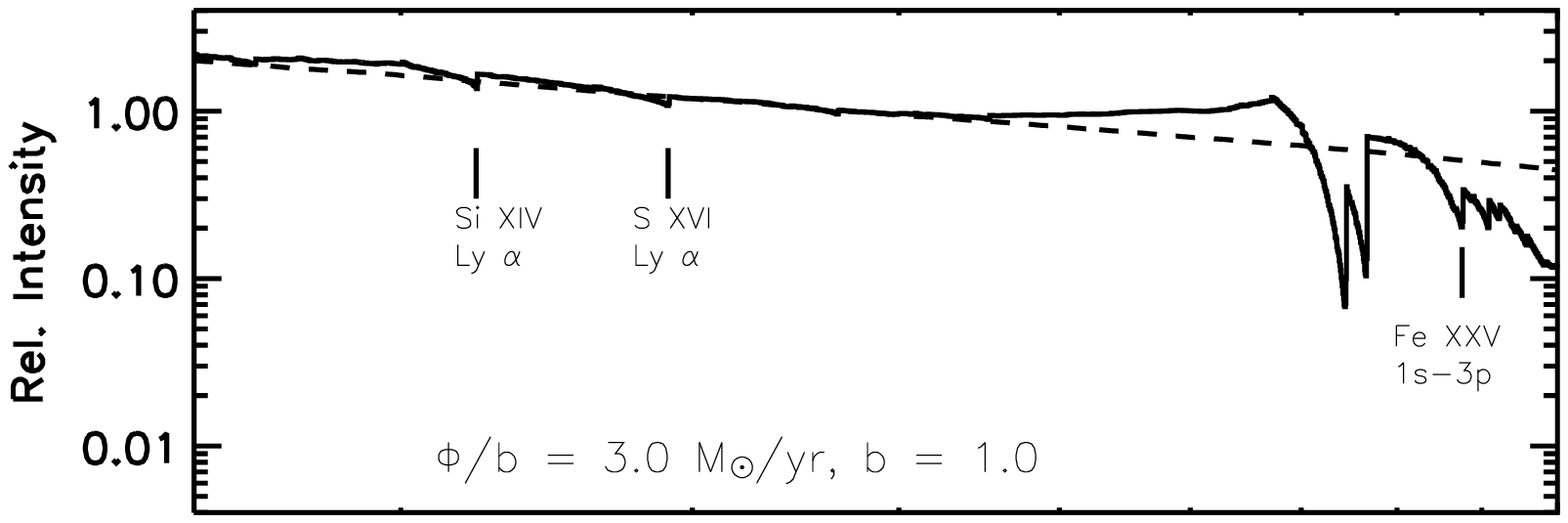, width=12cm,height=8cm}\\
\vspace*{-4.5cm}
\epsfig{file=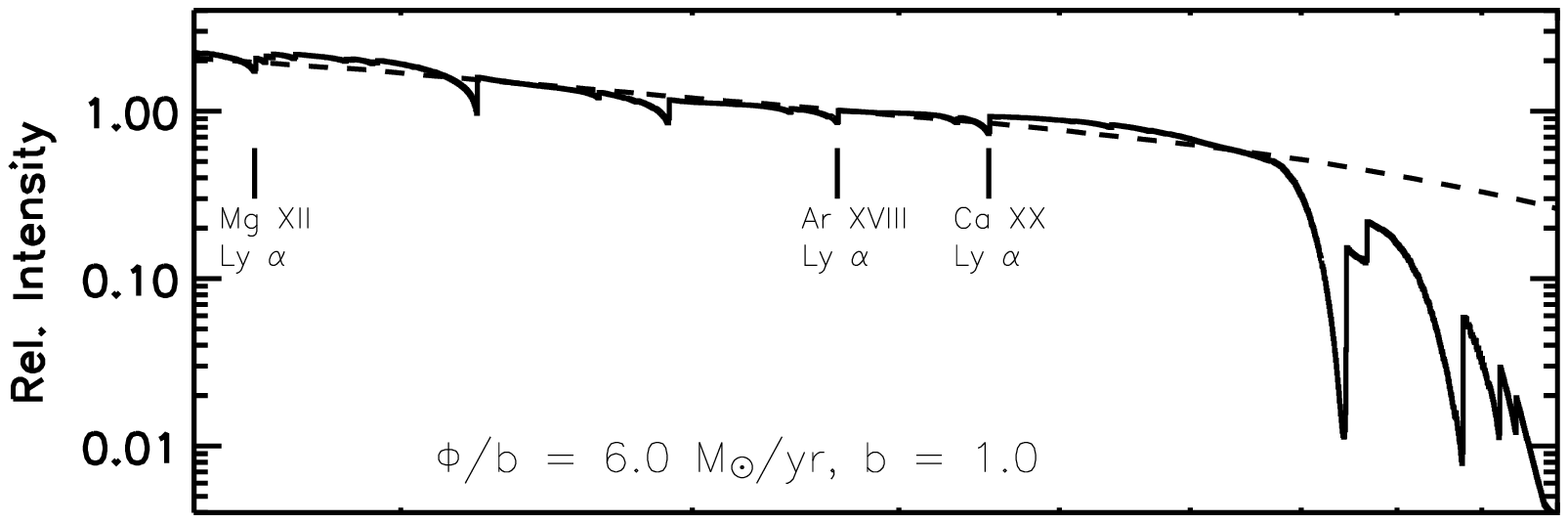, width=12cm,height=8cm}\\
\vspace*{-4.5cm}
\epsfig{file=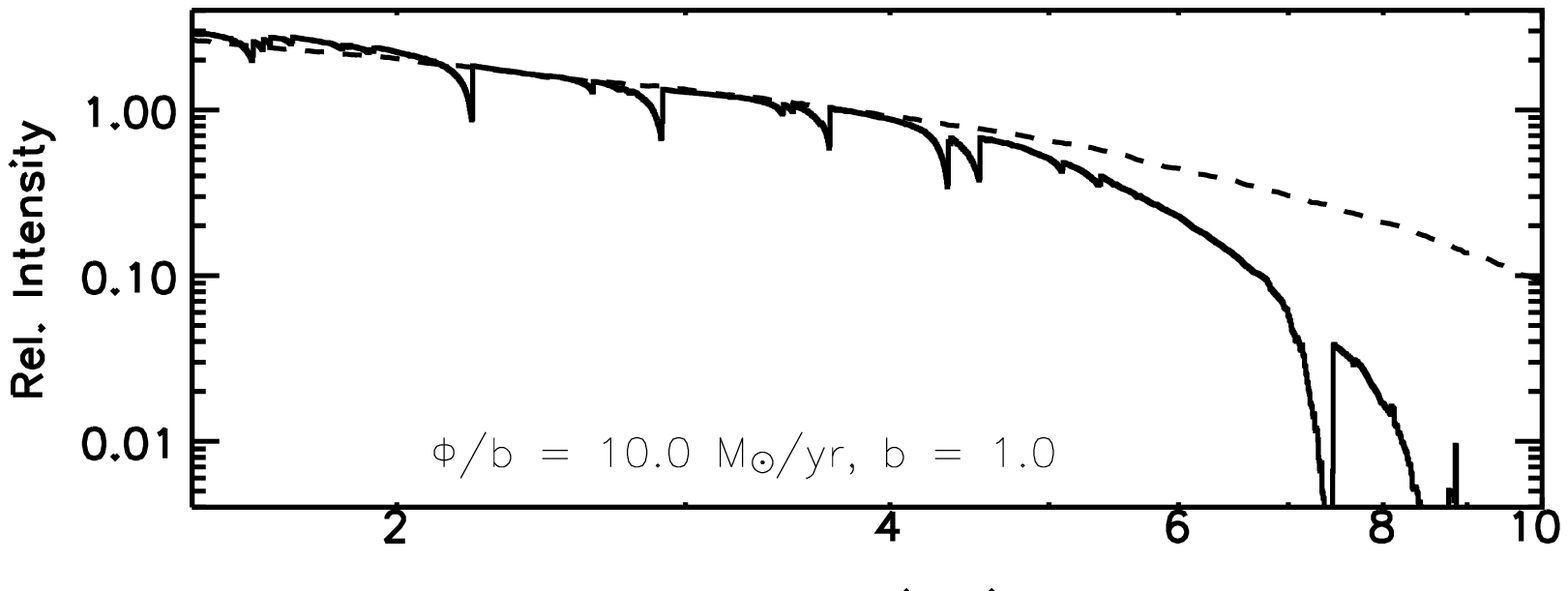, width=12cm,height=8cm}\\
\caption{2 -- 10 keV energy spectra computed from spherical ($b = 1$) models with 
(top to bottom)
$\Phi = 1$, 3, 6 and 
10 
M$_{\odot}$ yr$^{-1}$ (solid curves).
The energy is measured in the quasar rest-frame. 
In each case
$T_{e} = 1.3 \times 10^{6}$~K. The strongest spectral lines are identified
in the figure. 
The dashed lines show continuum spectra, computed 
in the absence of
bound-free processes and spectral lines.}
\end{figure*}

\section{Results}

\subsection{Spherical flow models}

Pounds et al. (2003a) suggest that the outflow 
from PG1211+143
may have a relatively wide opening angle and so it is instructive 
to begin with the consideration of the simplest wind geometry, that of spherical outflow 
(i.e. $\theta_{0} = 90$\textdegree, hence $b =1$).

\subsubsection{Model parameters}

In addition to the specification $\theta_{0} = 90$\textdegree, there are five parameters that must
be chosen: $v_{\infty}$, $R_{c}$, $\beta$, $T_{e}$ and $\Phi$ (see Section 2 for definitions of these
quantities).

The observed line shifts constrain $v_{\infty}$ directly. 
In all models discussed here, $v_{\infty} = 0.1c$ is adopted which is characteristic of 
the line shifts measured in both PG1211+143 and PG0844+349 by Pounds et al. (2003a,b).

It is reasonable to speculate that $R_{c}$ should be at least 
the radius of the last stable orbit around a
Schwarzchild black hole (i.e. $R_{c} > 3 R_{s}$ where $R_{s}$ is the 
Schwarzchild radius). 
Moreover, it is a general characteristic of 
winds in most astrophysical systems that the terminal velocity is comparable to the escape velocity in the region
from which the wind is launched. Given the choice $v_{\infty} = 0.1c$, this argument suggests 
$R_c = 100 R_{s}$ is appropriate, a value which is adopted throughout 
this investigation. 

In the absence of information on the probable shape of the velocity law in the flow, $\beta = 1/2$ is adopted in all
models. This value is theoretically attractive since it 
arises for the particular case that the driving force has the same radial dependence as
gravity and it is close to characteristic values for radiatively driven flows in other astrophysical 
objects (e.g. $\beta \sim 0.7$ in O stars [Groenewegen \& Lamers 1989]). 
There are objects for which significantly larger values (e.g.
$\beta \sim 3$ for Wolf-Rayet stars [see Ignace, Quigley \& Cassinelli 2003 and references therein]) 
have been suggested but values much smaller than $\beta = 1/2$ seem
improbable for radiatively driven flows.

As standard, $T_{e} = 1.3 \times 10^{6}$~K is adopted. This value is based on the primary black-body 
temperature fitted to the observed soft X-ray excess in PG1211+143 (Pounds et al. 2003a). The choice of
$T_{e}$ does not qualitatively affect the computed spectrum, but it does quantitatively influence
the equivalent widths of spectral features. In principle, if all important heating and cooling processes
were included in the calculation, $T_{e}$ could be computed rather than assumed. However, this goes beyond
the scope of this paper in which the primary objective is spectral synthesis. 
To indicate the sensitivity of the results presented here to $T_{e}$, the Appendix briefly discusses 
models with temperatures both higher and lower than the standard temperature adopted here.

The mass-loss rate, $\Phi$ is one of the primary unknowns in this investigation. Based on simple fits to the 
{\it XMM-Newton}
data, Pounds et al. (2003a) suggest a mass-loss rate $\Phi \sim 3$~M$_{\odot}$ yr$^{-1}$. For this investigation, 
calculations have been carried out for spherical models
with $\Phi = 1$, 3, 6, and 10~M$_{\odot}$ yr$^{-1}$. 
The results of these calculations are presented in the next section.

\begin{table*}
\caption{Equivalent width (EW) and full-width at half-maximum (FWHM) for the 
computed absorption components
of the
S~{\sc xvi} and Mg~{\sc xii} Lyman~$\alpha$ lines and the combined Fe~{\sc xxvi}/{\sc xxv}
feature at 7.5~keV for spherical wind models ($\theta_{0} = 90$\textdegree, $b=1$). 
The last row in the table gives the observational constraints taken from 
Pounds et al. (2003a).
The last column of the table gives the radius in the outflow at which the
electron scattering optical depth $\tau_{e} = 1$ in the radial direction.
The Monte Carlo simulations are accurate to about $\pm 10$ per cent in EW 
and $\pm 300$~km~s$^{-1}$ in FWHM.}
\begin{tabular}{lccccccc} \hline
{Model} & \multicolumn{2}{c}{Fe (7.5 keV)} & 
\multicolumn{2}{c}{S~{\sc xvi} Lyman~$\alpha$} &
\multicolumn{2}{c}{Mg~{\sc xii} Lyman~$\alpha$} & $r(\tau_e = 1)$\\
$\Phi / b$ & EW & FWHM & EW & FWHM & \\
(M$_{\odot}$ yr$^{-1}$) & (eV) & (km~s$^{-1})$ 
& (eV) & (km~s$^{-1}$) & (eV) & (km~s$^{-1}$) &($R_{s}$) \\ \hline
1.0 & 140 & 21,000$^a$ & --$^b$ & --$^b$ & --$^b$ & --$^b$ & 100\\
3.0 & 360 & 21,000$^a$ & 10 &  7,400  & --$^b$ & --$^b$ &200  \\
6.0 & 380 & 20,000$^a$ & 19 &  5,300  & 6 & 4,800 & 370\\
10.0& 280 & 16,000$^a$ & 30 &  4,500  & 8 & 4,000 & 590\\
{Observations} & $95 \pm 20$ & $\sim 12,000$ & $32 \pm 12$ & &
$15 \pm 6$ & \\\hline
\end{tabular}

\noindent $^a$ To obtain a single FWHM for the Fe blend at 7.5~keV, the computed spectrum was
convolved with a Gaussian of FWHM = 10,000~km~s$^{-1}$. The FWHM reported in the table is that
measured from the convolved spectrum.

\noindent $^b$ In some cases, several of the lines do not appear 
or are too weak for reliable calculation of their profiles above the
Monte Carlo noise.

\end{table*}

\subsubsection{Computed spectra and discussion}

Figure~3 shows spectra computed in the 2 -- 10 keV region from models
with the standard parameters discussed above and the four values of $\Phi$ under
consideration (1, 3, 6 and 10~M$_{\odot}$ yr$^{-1}$). In each case, the continuum level as computed in a separate run of the code in which
only Compton scattering by electrons is considered (i.e. no bound-free continua or lines are 
included) is plotted for comparison.

These model spectra are rather encouraging: in all four cases they are 
consistent with the observations in that the strongest absorption feature 
is, by a considerable margin, due to the Lyman $\alpha$ line of hydrogen-like
iron (Fe~{\sc xxvi}) and the 1s -- 2p resonance line of helium-like iron
(Fe~{\sc xxv}). Comparing the four spectra plotted in Figure~3 shows
that as the mass-loss rate is increased the Fe~{\sc xxv} line becomes 
stronger relative to the Fe~{\sc xxvi} line. This is a consequence
of the response of the iron ionization balance to the 
density -- higher densities favour formation of the lower ionization
species. The ionization balance is discussed further in 
Section 5.2.

At the lowest densities considered ($\Phi = 1$~M$_{\odot}$ yr$^{-1}$) the
blended iron feature is the only strong feature in the spectrum. At higher
densities weak lines appear at lower energies, most importantly the
Lyman $\alpha$ lines of hydrogen-like silicon and sulphur. The sulphur line is
observed in PG1211+143 but the region around the silicon line is marred by
instrumental effects in the data preventing a firm identification (Pounds et al.
2003a).
The Lyman $\alpha$ line of hydrogen-like magnesium is also present in the data.
The high density models do predict this feature but it is always weak.

Various absorption features appear at
energies above $\sim 8.5$~keV and become stronger as the
density is increased. Most prominent of the high energy ($> 8$~keV) 
absorption features is the 1s -- 3p resonance line
of Fe~{\sc xxv} but there are several other contributers to the absorption 
including the Fe~{\sc xxvi} Lyman~$\beta$ line (at around 9 keV), the
ground state photoionization edges of Fe~{\sc xxv} and {\sc xxvi} and lines
of Ni~{\sc xxvii} and {\sc xxviii}.
In the observations of PG1211+143 (Pounds et al. 2003a) an absorption line
is detected at $\sim 8.7$~keV but it is less statistically significant than the
stronger feature at $\sim 7.5$~keV.

In addition to blueshifted absorption there is some weak 
redshifted emission associated with several of the features. This appears in
the spherical models because there is no obstruction to observing the receding
back hemisphere of the flow. These emission
features are largely absent from the spectra computed from bi-conical 
models (Section 5.2) in which the receding flow is obscured.

For quantitative comparison, Table~2 gives the equivalent width and full-width
at half-maximum (FWHM) of the iron feature at 7.5 keV and the observed S~{\sc xvi} and 
Mg~{\sc xii} Lyman~$\alpha$
lines for the model spectra. 
None of the spherical models predict strong, sharp absorption lines due to
lower ionization elements at softer energies (e.g. the O~{\sc viii} Lyman~$\alpha$ line).
Since the Fe~{\sc xxvi} and 
Fe~{\sc xxv} components of the 7.5 keV feature are not separately identified in
the data, these two lines are considered as one blended feature. 
To determine a single FWHM for this feature, the
computed spectra were convolved with a Gaussian of FWHM = 10,000~km~s$^{-1}$ to
simulate the {\it XMM-Newton} resolution. The tabulated width is that obtained from the
convolved spectrum.
For completeness, the table also gives the radial position in the flow at which the optical depth due to 
electron scattering ($\tau_e$) is unity in the radial direction (this is the definition used for the 
``photospheric'' radius by Pounds et al. [2003a] and King \& Pounds [2003]).

\begin{figure*}
\vspace*{-4.5cm}
\epsfig{file=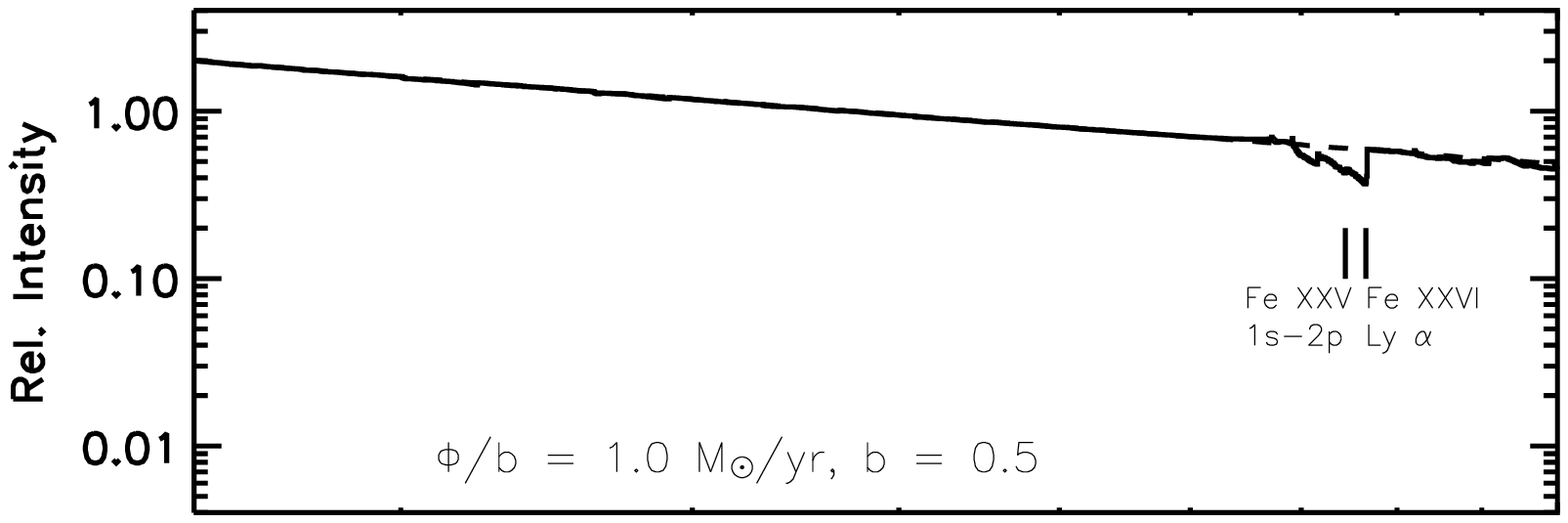, width=12cm, height=8cm}\\
\vspace*{-4.5cm}
\epsfig{file=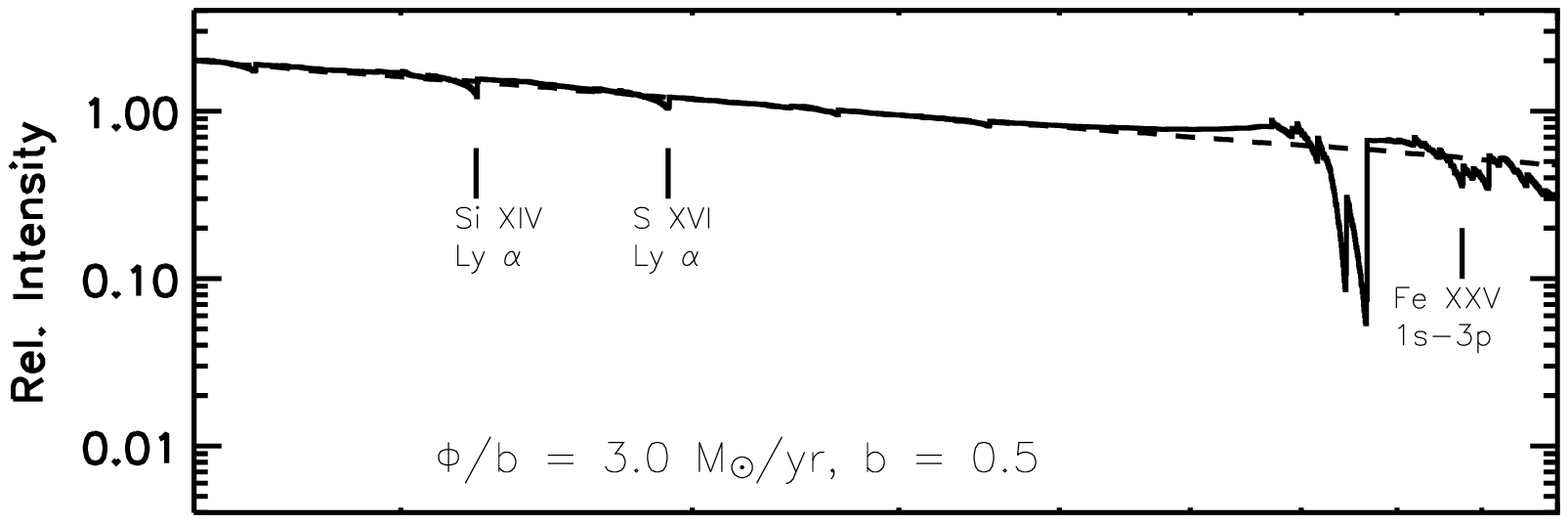, width=12cm,height=8cm}\\
\vspace*{-4.5cm}
\epsfig{file=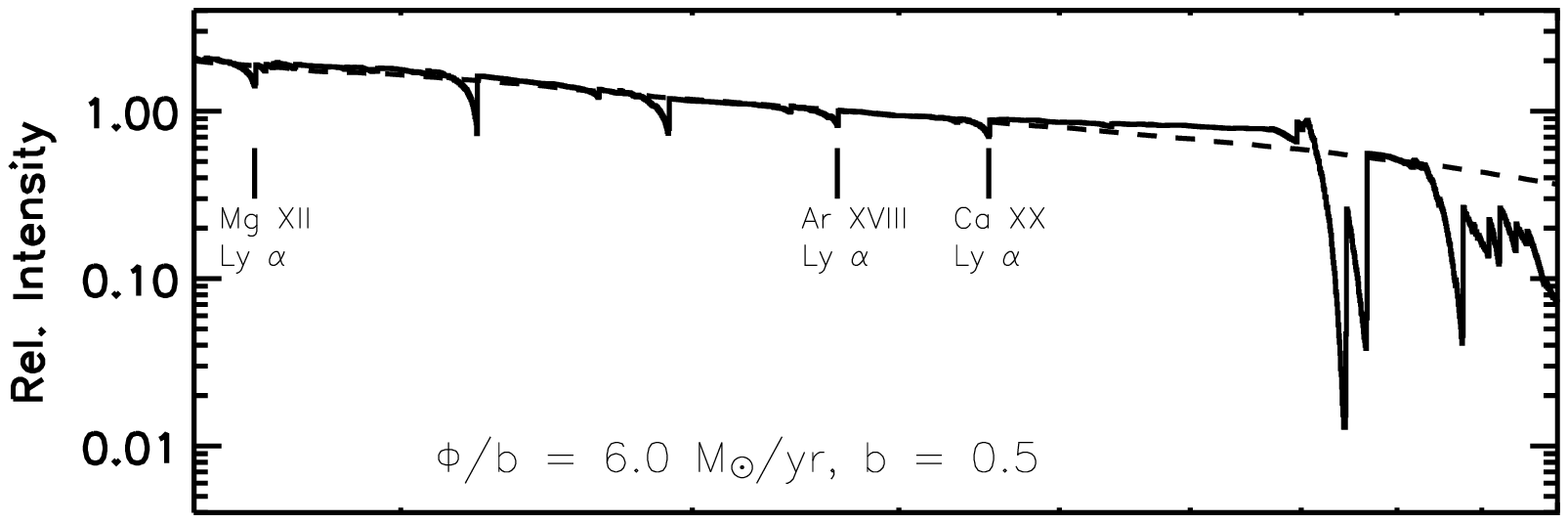, width=12cm,height=8cm}\\
\vspace*{-4.5cm}
\epsfig{file=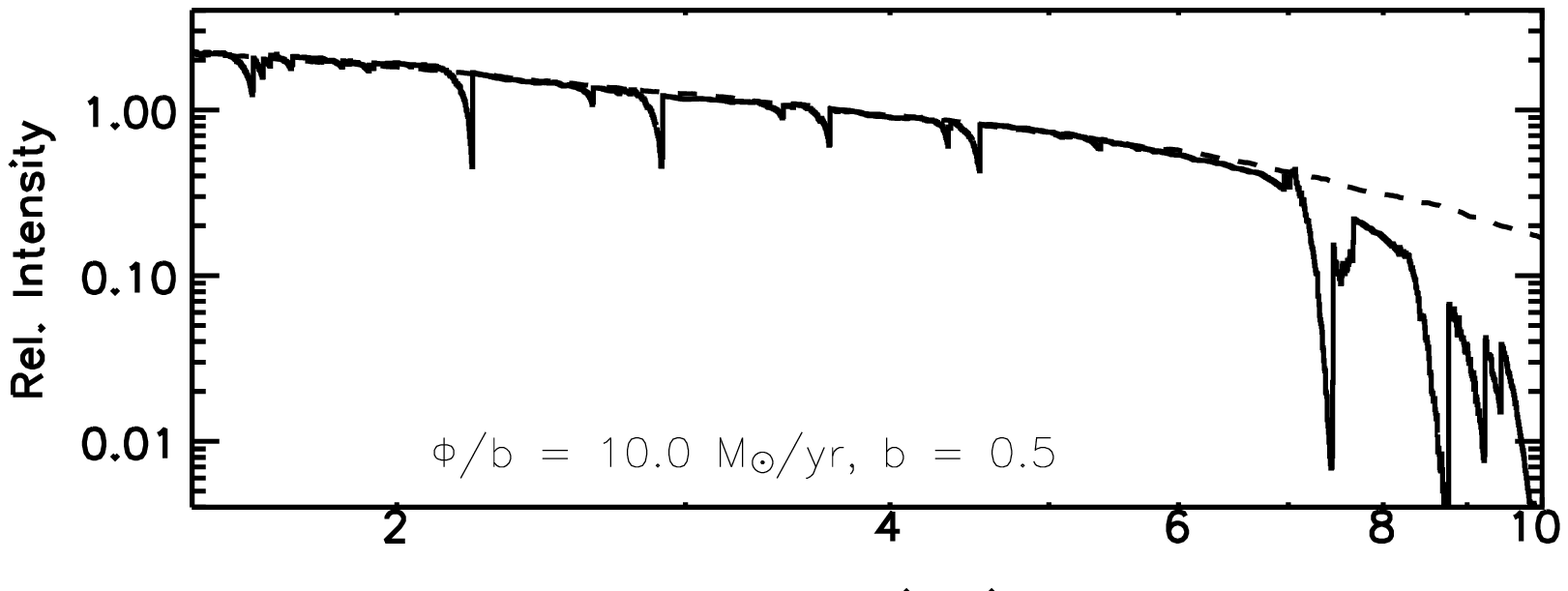, width=12cm,height=8cm}\\
\caption{
As Figure~3 but showing results for bi-conical models with $\theta_0 = 60$\textdegree~($b = 0.5$).
}
\end{figure*}

Figure~3 and Table~2 show
that the sulphur and magnesium lines become stronger and narrower as the 
mass-loss rate is increased. 
It is apparent from Figure~3 that as the density is increased the
two components of the iron 7.5~keV feature also become narrower.
However, 
the width of the combined feature (see Table~2) does not vary much
until the density is high enough that only the Fe~{\sc xxv} component
is strong. 

Although encouraging for the reasons discussed above, the spherical 
models have 
problems when confronted by the data. 
The high mass-loss rate models
($\Phi = 6.0$ and 10~M$_{\odot}$ yr$^{-1}$) can be easily ruled out based 
on their predicted
spectra at $>8$~keV: these models favour a rapid decline in hard X-ray flux
as a function of energy which is not supported by the observations. 
This predicted decline occurs as a result of both multiple Compton scattering
and absorption by spectral lines and continua.

In the lower mass-loss rate models, the 
7.5 keV iron feature is slightly too strong while the 
sulphur and magnesium Lyman~$\alpha$ lines are
too weak or absent. 
Also, these models do not predict significant absorption by the 
Lyman~$\alpha$ lines of hydrogen-like carbon, nitrogen or oxygen.
These lines have been detected in the RGS spectrum of PG1211+143
(Pounds et al. 2003a) and have similar
blueshifts to the hard X-ray lines, suggesting that they may form in the
same outflow.

In addition, the linewidths of the features in the low density models
are likely to be too large.
The 7.5~keV iron feature is unresolved in the 
data, implying 
that its width should be $\leq 12,000$~km~s$^{-1}$ (Pounds et al. 2003a).
For all but the highest density model, the predicted width of the feature
at 7.5~keV is noticeably greater than this. It is important to note, 
however, that the widths of the two lines individually (Fe~{\sc xxvi} 
Lyman~$\alpha$ and Fe~{\sc xxv} 1s--2p) are significantly smaller than 
that of the blended feature. Therefore if the observed line corresponds
to only one of the computed K~$\alpha$
features (as suggested by Pounds et al. 2003a) the widths are compatible.
The sulphur and magnesium lines are predicted to be significantly narrower 
than the iron line. Constraints on the widths of these lines are not 
provided by the data. However, 
profiles of lines arising from lower ionization ions in the soft X-ray RGS 
spectrum suggest linewidths 
$\leq 2,000$~km~s$^{-1}$ (Pounds et al. 2003a). The widths of the
soft X-ray lines do not place direct constraints on the properties required of
hard X-ray spectral features but if the sulphur and magnesium Lyman $\alpha$ lines form
under similar conditions to lines observed at softer energies (e.g. the 
observed O~{\sc viii} Lyman~$\alpha$ line)
it is to be expected that their linewidths should be comparable. This is weak
additional evidence against the low density spherical models.

Given the quality of the data and the various model
uncertainties (including the element abundances) it is unreasonable to
expect very good quantitative agreement 
but it is worthwhile to search for model conditions
which might reproduce the observed properties of the spectrum
more accurately.
In the next section it is shown that improvements 
can be obtained by lifting the assumption of sphericity 
and considering bi-conical flows such as discussed by Pounds et al. (2003a) and
King \& Pounds (2003).

\subsection{Bi-conical flow models}

\begin{figure*}
\vspace*{-4.5cm}
\epsfig{file=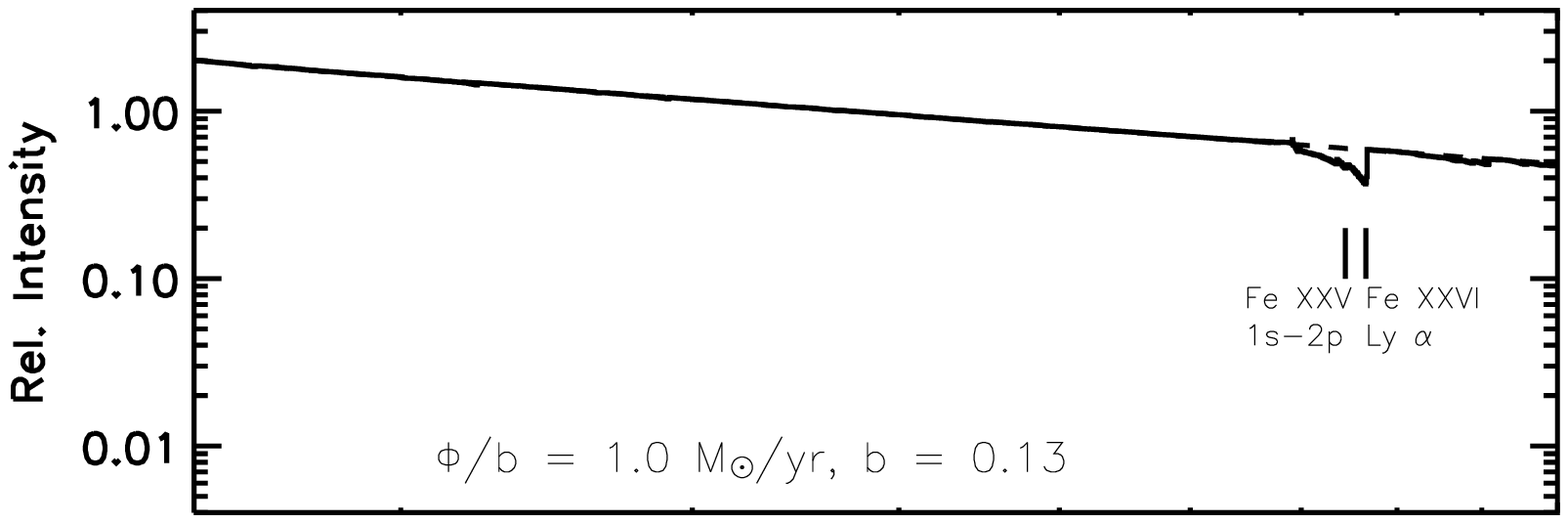, width=12cm, height=8cm}\\
\vspace*{-4.5cm}
\epsfig{file=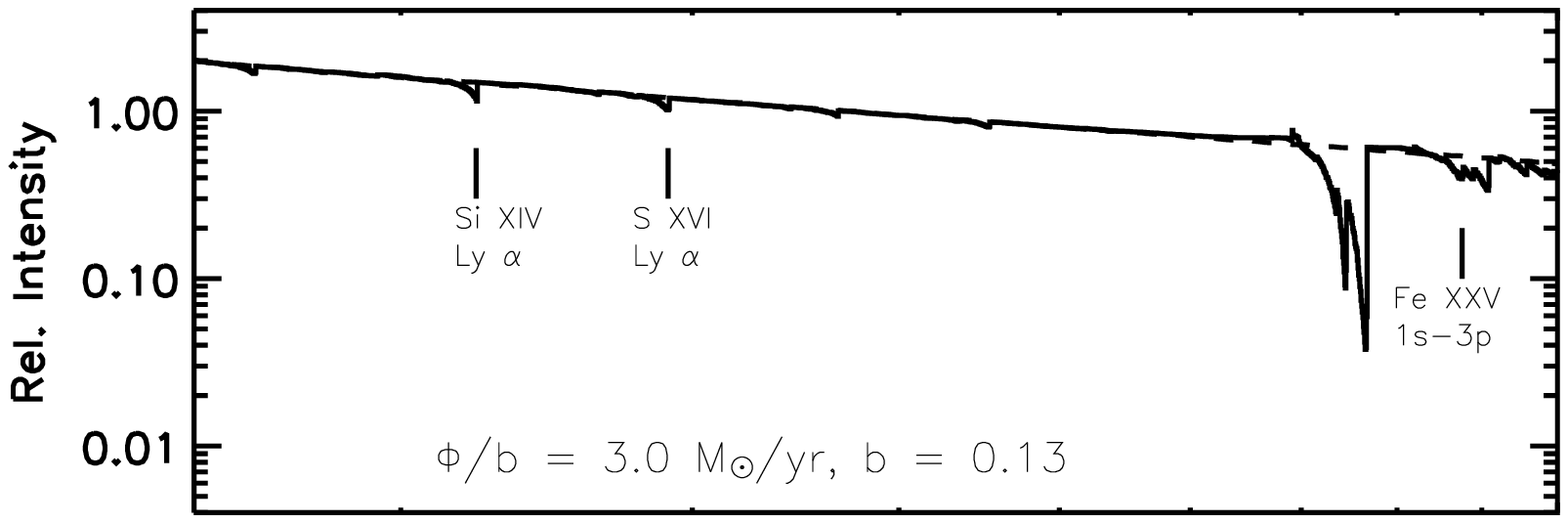, width=12cm,height=8cm}\\
\vspace*{-4.5cm}
\epsfig{file=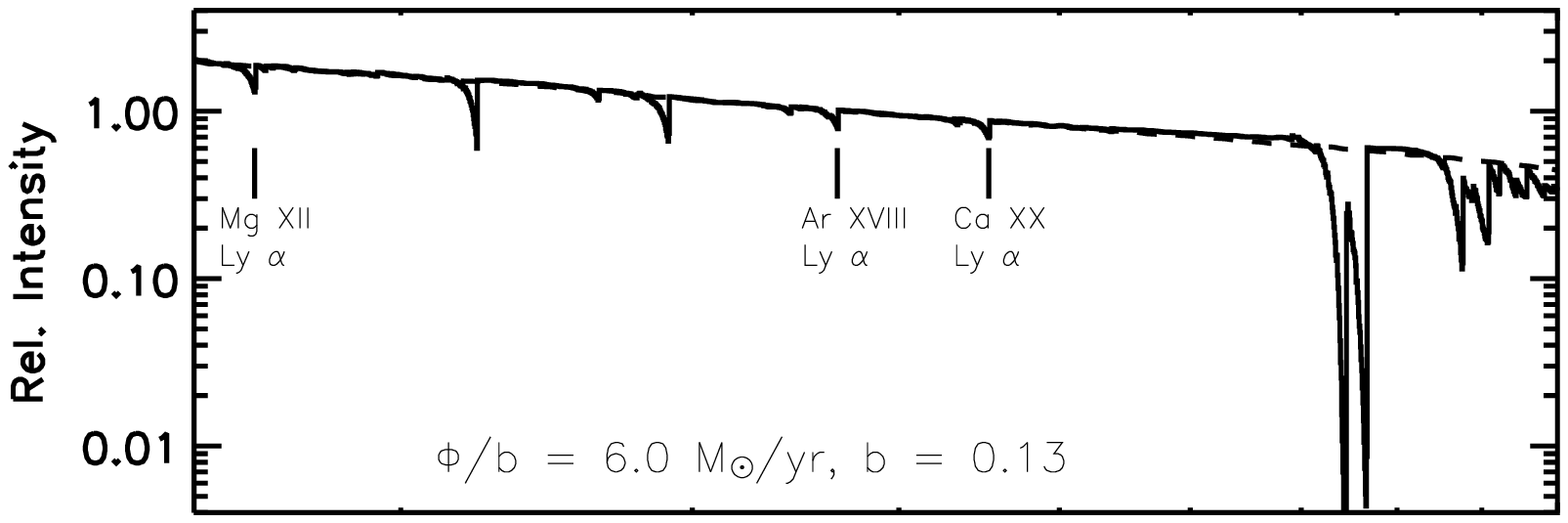, width=12cm,height=8cm}\\
\vspace*{-4.5cm}
\epsfig{file=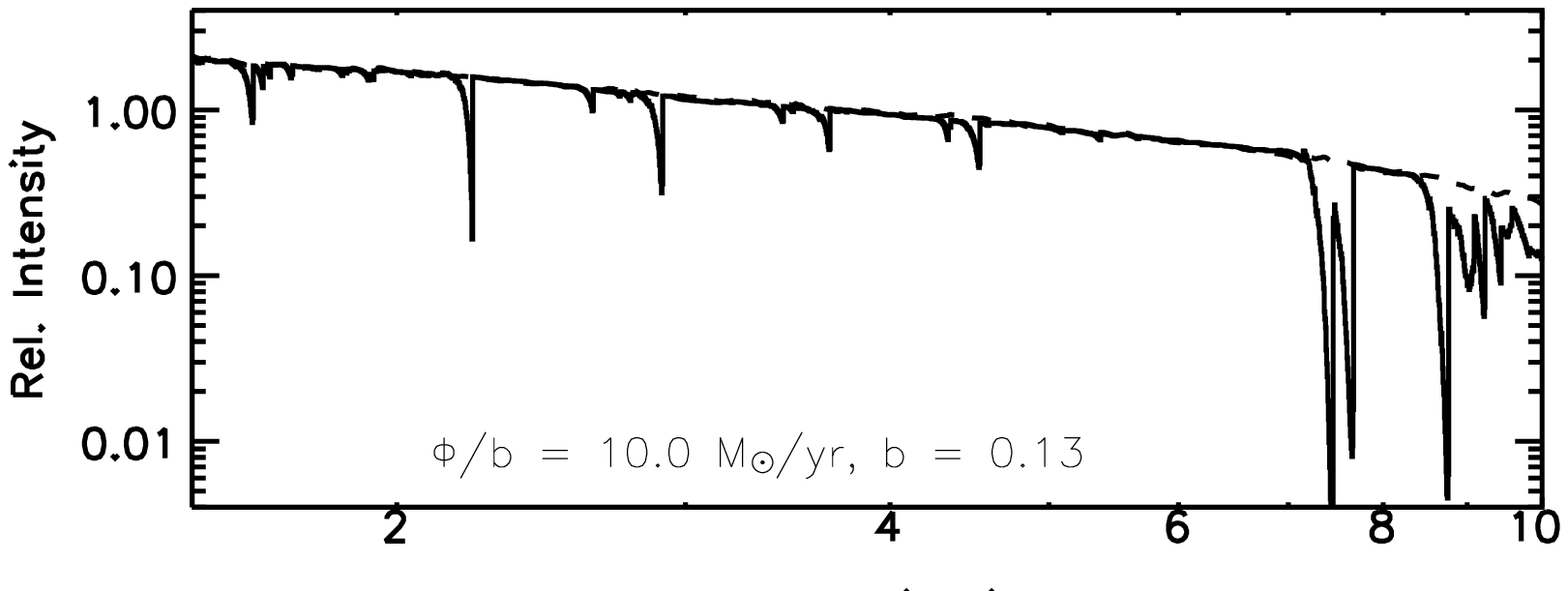, width=12cm,height=8cm}\\
\caption{
As Figure~3 but showing results for bi-conical models with $\theta_0 = 30$\textdegree~($b = 0.13$).
}
\end{figure*}

Models in which $\theta_{0} < 90$\textdegree~($b < 1$) are now considered and 
contrasted with the spherical models discussed above. 

\subsubsection{Model parameters}
Throughout this
section the parameters $v_{\infty}$, $R_{c}$, $\beta$ and $T_{e}$ remain
fixed at the values used for the spherical models. Spectra are 
computed for models with
$\theta_{0} = 60$\textdegree~ ($b = 0.5$) and $\theta_{0} = 30$\textdegree~
($b = 0.13$). For both values
of $\theta_{0}$, models with four values of $\Phi$ have been 
considered. For ease of comparison between the models, the values of $\Phi$
adopted are chosen to give the same densities as in the
spherical models rather than the same total mass-loss rates (i.e. the
values of $\Phi$ considered are such that
$\Phi / b = 1$, 3, 6 and 10~M$_{\odot}$ yr$^{-1}$).

\subsubsection{Computed spectra and discussion}

Figures 4 and 5
show the spectra computed from models with $\theta_0 = 60$\textdegree~and 30\textdegree~
respectively. Individually, Figures~4 and 5 show similar trends in the spectra
as a function of density as the spherical models in Section 5.1.

More interestingly,
by comparing Figures~3, 4 and 5,
it is apparent that as $\theta_{0}$
is reduced, the iron feature at 7.5~keV remains strong
but both components of the absorption
become sharper and
the relative strength of the Fe~{\sc xxv}
and {\sc xxvi} features varies with $\theta_{0}$ such that the higher
ionization line is strongest when the opening angle is smallest.

\begin{figure*}
\vspace*{-0.6cm}
\hspace{-0.5cm}
\epsfig{file=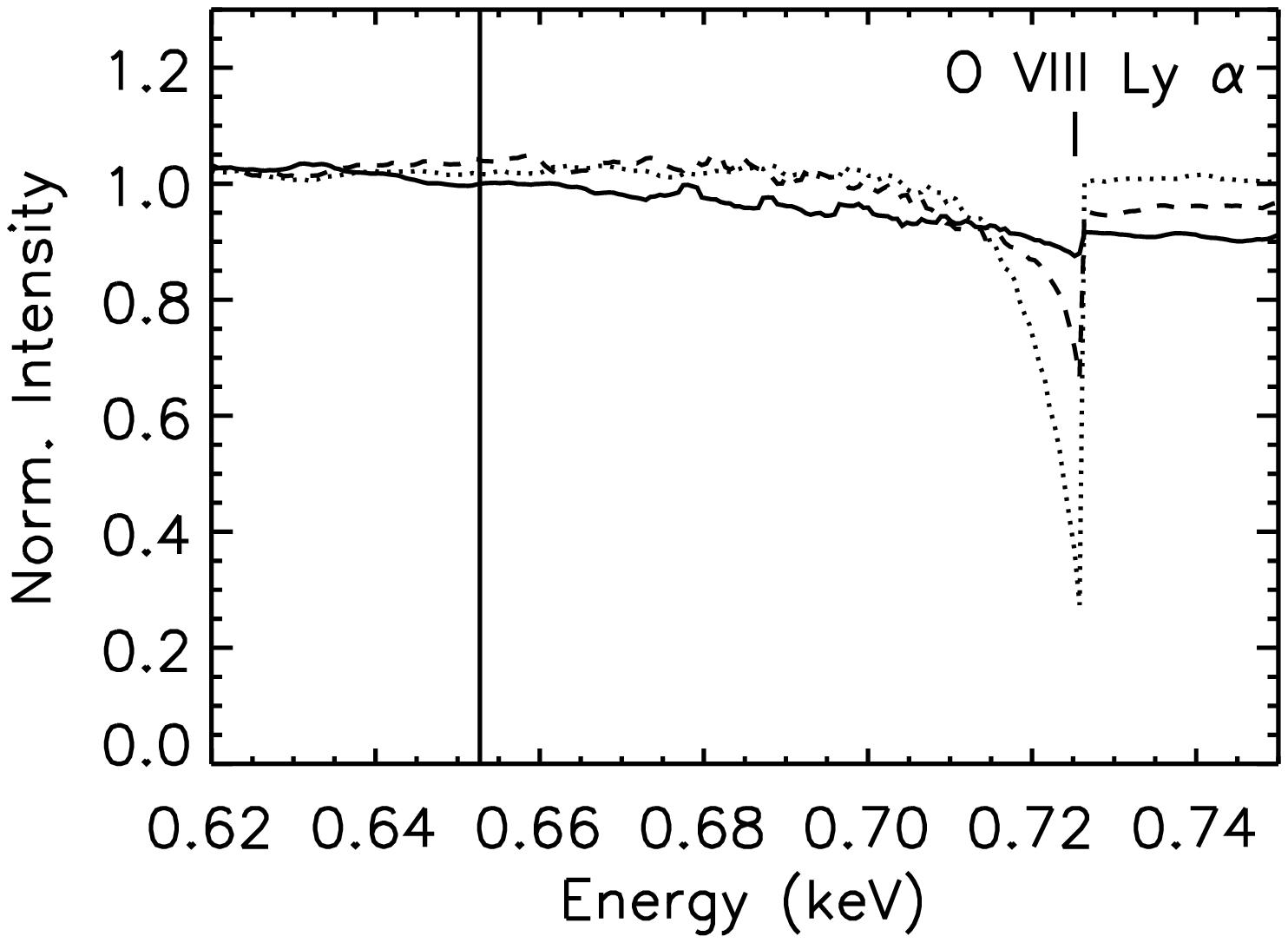, height = 5cm, width = 5cm}
\hspace{-0.5cm}
\epsfig{file=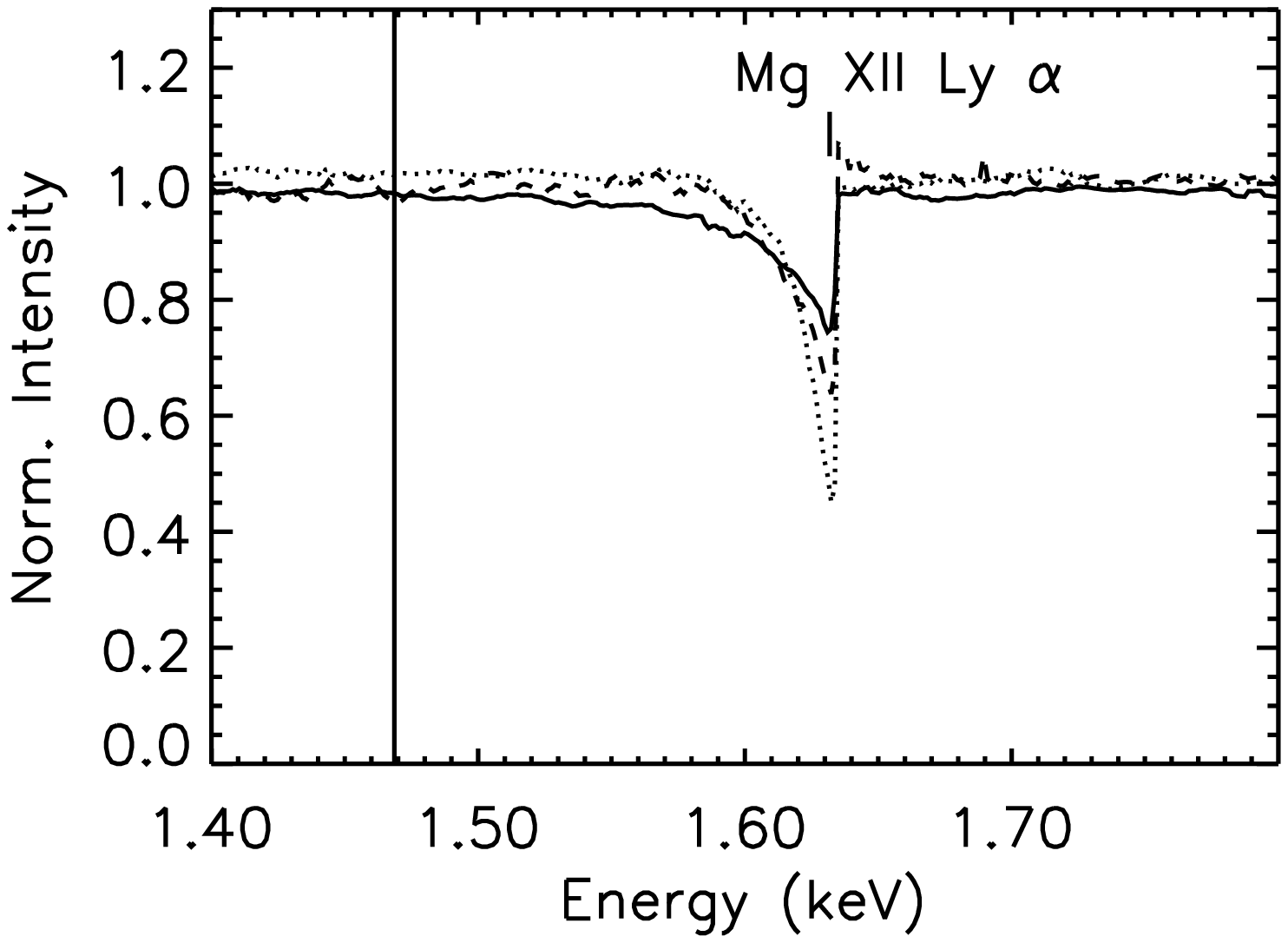, height = 5cm, width = 5cm}
\hspace{-0.5cm}
\epsfig{file=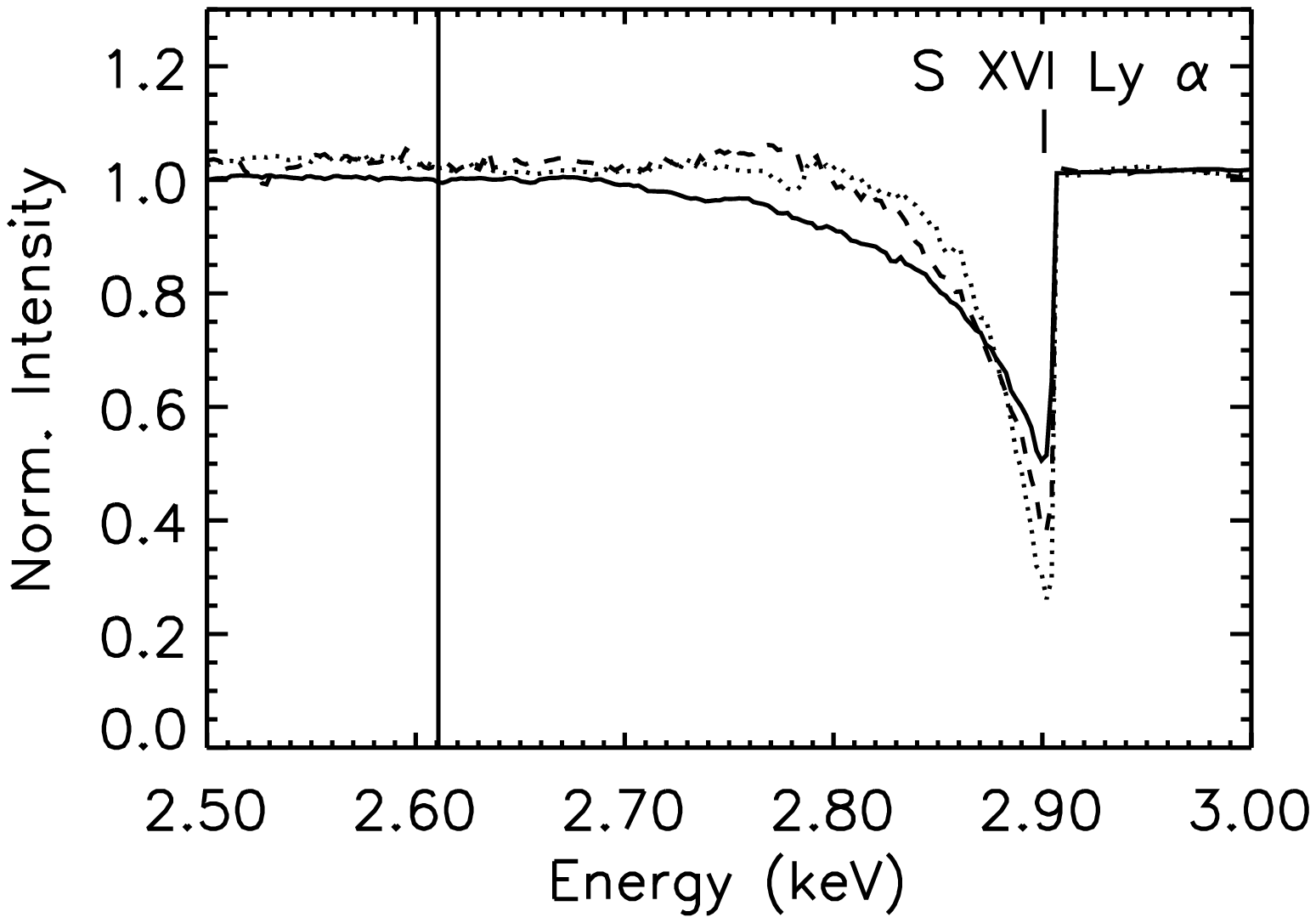, height = 5cm, width = 5cm}
\caption{Detail of the computed 
O~{\sc viii}, Mg ~{\sc xii} and S~{\sc xvi} Lyman $\alpha$ line profiles 
for the highest density models ($\Phi/b = 10$~M$_{\odot}$ yr$^{-1}$).
The spectra are normalised to the continuum.
In each panel, 
the solid curve shows results for $\theta_{0} = 90$\textdegree~(spherical
flow), the dashed curve for 
$\theta_{0} = 60$\textdegree~($b= 0.5$),
and
the dotted curve for
$\theta_{0} = 30$\textdegree~($b= 0.13$).
The labels identify the line positions at the terminal velocity
of the outflow.
The solid vertical
lines indicate the rest energies of the lines. 
The small scale structure that can be seen at the level of $\sim 5$ per cent
is Monte Carlo noise in the simulations.}
\end{figure*}

The opening angle has a more significant effect on the weaker, lower
energy lines.
To illustrate 
this influence
Figure 6
compares the S~{\sc xvi}, Mg~{\sc xii} and O~{\sc viii} Lyman~$\alpha$ line profiles 
from the highest density models ($\Phi/b = 10$~~M$_{\odot}$ yr$^{-1}$) for different
values of $\theta_{0}$ in detail.
The calculations reveal an interesting trend: as $\theta_{0}$ is reduced these lines
become narrower and deeper, bringing them into significantly closer agreement with
the constraints placed on the line profiles by the observations.
The rapidity of this effect is inversely
correlated with the ionization potential of the species:
the relatively low ionization O~{\sc viii} Lyman~$\alpha$ line at 
0.72~keV, which is the strongest of the lines of various hydrogen-like
elements identified by Pounds et. al (2003a) in their RGS data,
shows more dramatic variations than the intermediate ionization
Mg~{\sc xii} and S~{\sc xvi} lines. 

The dependence of the spectral features on both $\theta_{0}$ and $\Phi / b$
can be readily understood in terms of the influence of these model parameters
on the computed ionization structure of the wind. To illustrate this, Figure~7
shows the ionization fractions of S~{\sc xvi} and S~{\sc xv} in four models
(combinations of $\theta_{0} = 90$\textdegree~[spherical] or 30\textdegree~with
$\Phi / b = 1.0$ or 6.0~M$_{\odot}$ yr$^{-1}$). 

\begin{figure*}
\vspace*{-0.6cm}
\hspace{-0.5cm}
\epsfig{file=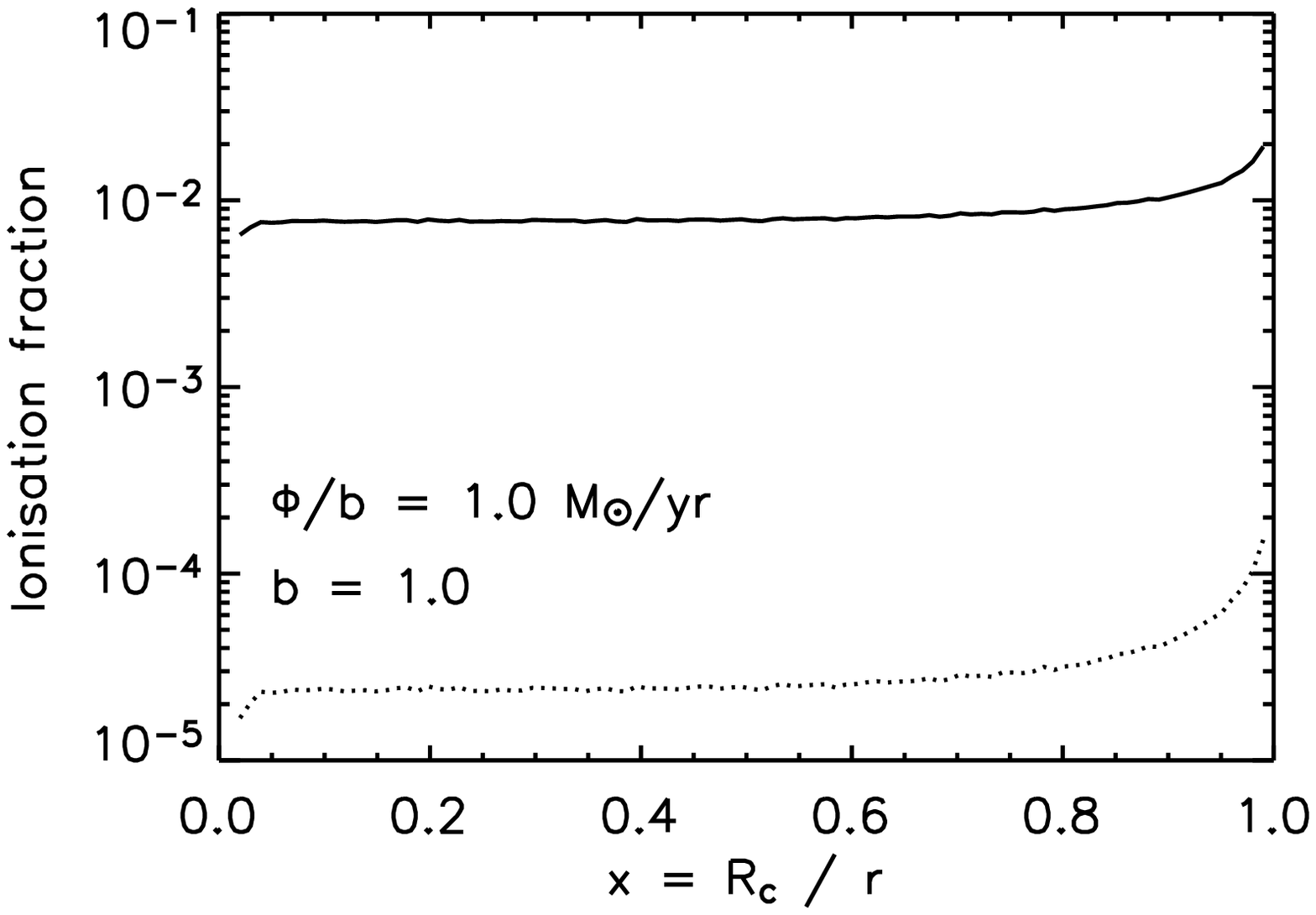, height = 5cm, width = 5cm}
\hspace{-0.5cm}
\epsfig{file=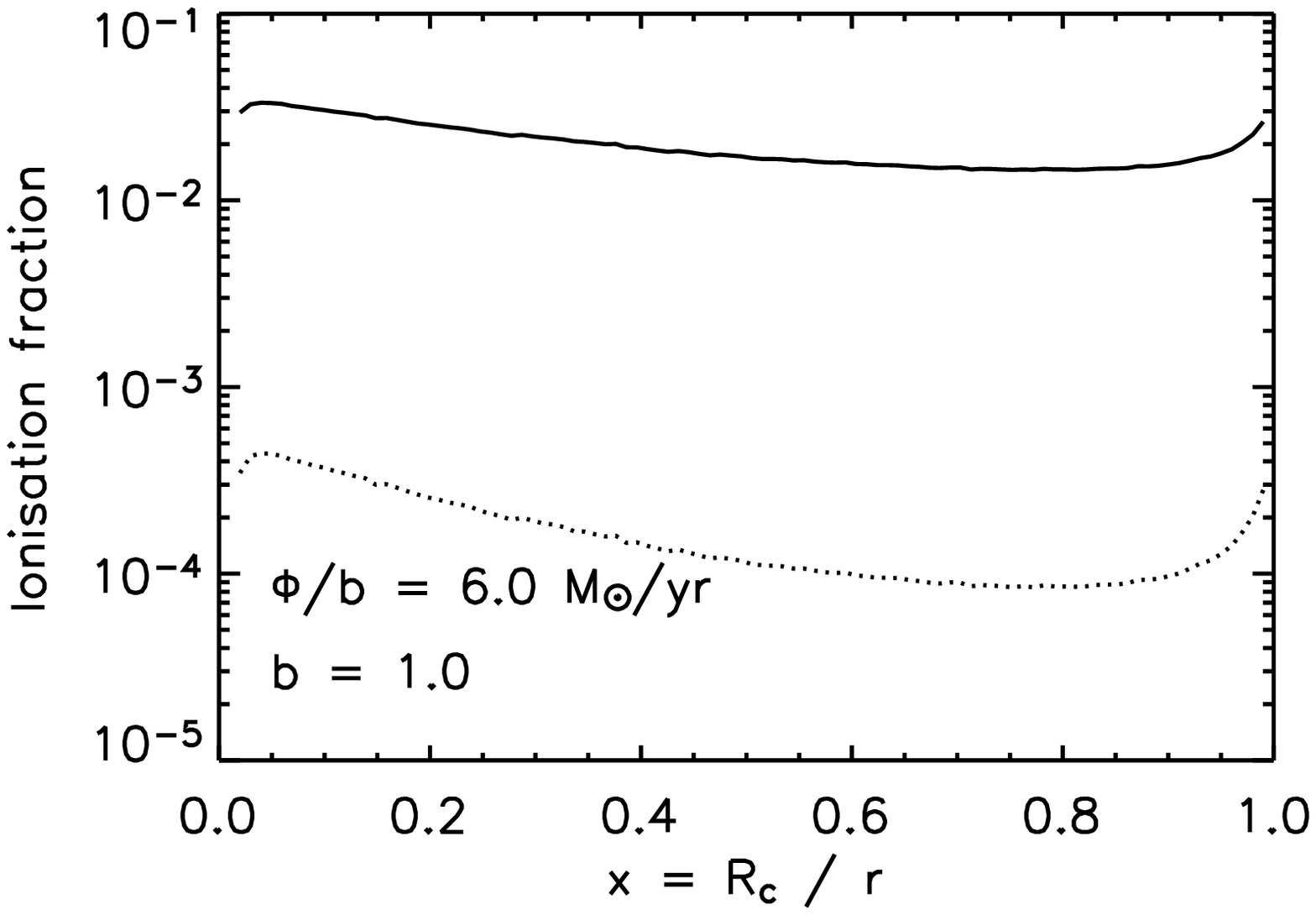, height = 5cm, width = 5cm}\\
\vspace*{-0.6cm}
\hspace{-0.5cm}
\epsfig{file=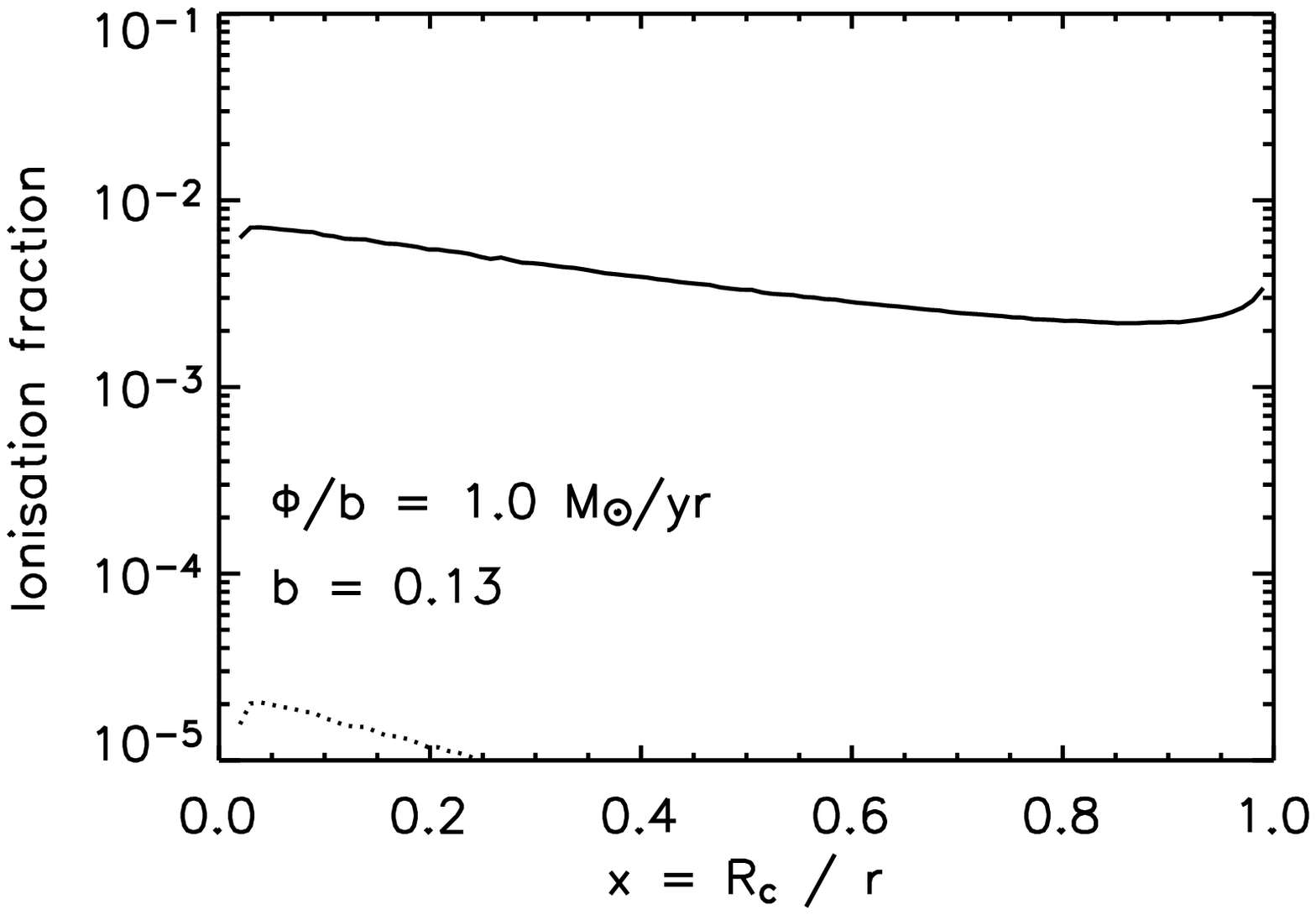, height = 5cm, width = 5cm}
\hspace{-0.5cm}
\epsfig{file=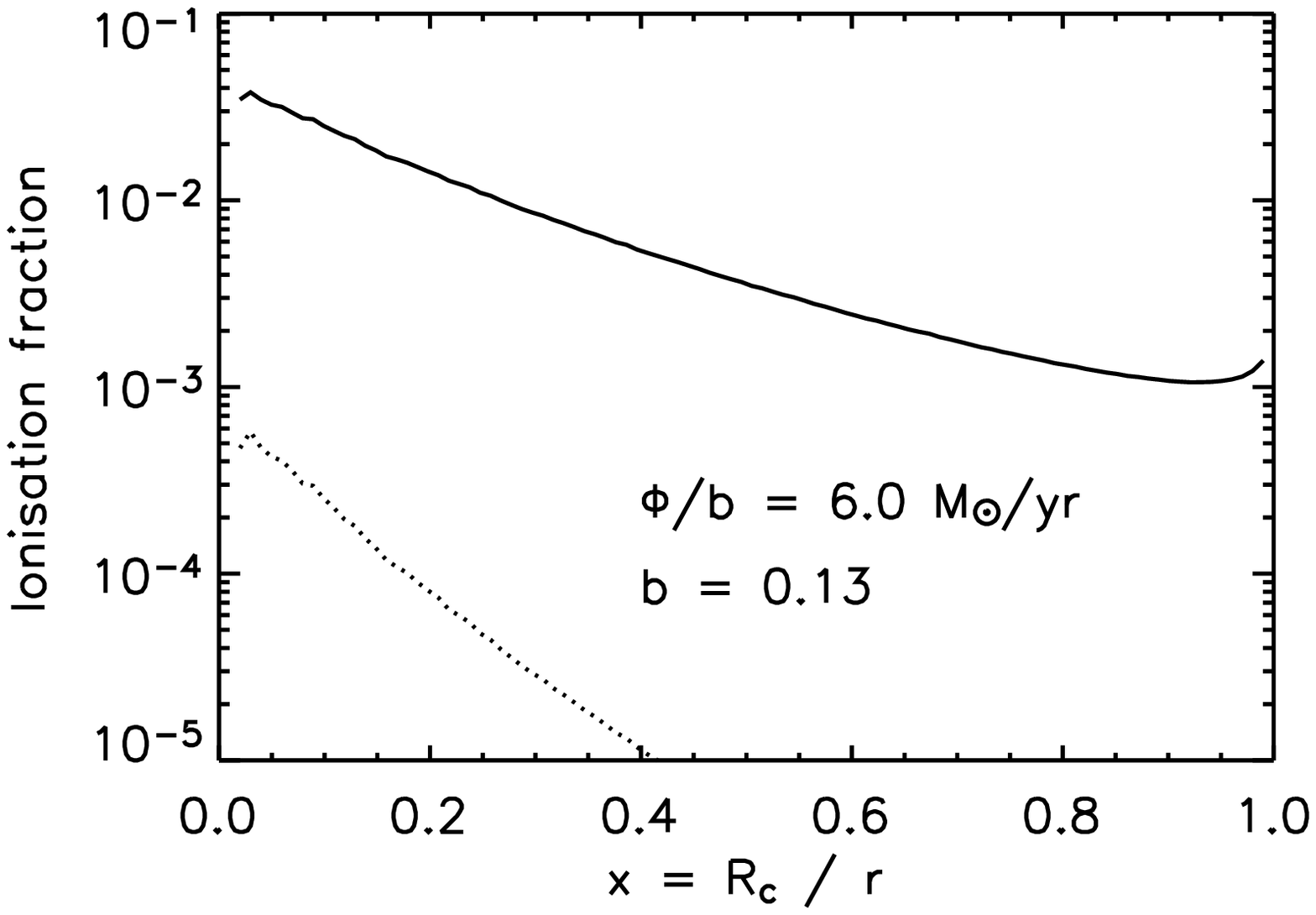, height = 5cm, width = 5cm}\\
\caption{
Ionization fractions of S~{\sc xvi} (solid
curves) and {\sc xv} (dotted curves) as a function of $x = R_{c} / r$
where $r$ is the radial distance from the central black hole. The four panels
correspond to models with $\Phi / b = 1$~M$_{\odot}$ yr$^{-1}$
and $\theta_{0} = 90$\textdegree
~ 
(upper left), 
$\Phi / b = 6$~M$_{\odot}$ yr$^{-1}$ and $\theta_{0} = 90$\textdegree~
(upper right),
$\Phi / b = 1$~M$_{\odot}$ yr$^{-1}$ and $\theta_{0} = 30$\textdegree~
(lower left) and
$\Phi / b = 6$~M$_{\odot}$ yr$^{-1}$ and $\theta_{0} = 30$\textdegree~
(lower right).
Ionization fractions of
oxygen, magnesium and iron behave in a similar manner to those of sulphur.
}
\end{figure*}

\begin{table*}
\caption{Equivalent width (EW) and full-width at half-maximum (FWHM) for the 
computed
S~{\sc xvi}, Mg~{\sc xii} and 
O~{\sc viii} Lyman~$\alpha$ lines and the combined Fe~{\sc xxvi}/{\sc xxv}
feature at 7.5~keV for bi-conical wind models.
For most strong lines, 
the Monte Carlo simulations are accurate to about $\pm 10$ per cent in EW 
and $\pm 300$~km~s$^{-1}$ in FWHM. The uncertainties are larger for weaker lines.
The last row in the table gives the observational constraints taken from 
Pounds et al. (2003a).
}
\begin{tabular}{lccccccccc} \hline
\multicolumn{2}{c}{Model} & \multicolumn{2}{c}{Fe (7.5 keV)} & 
\multicolumn{2}{c}{S~{\sc xvi} Lyman~$\alpha$} &
\multicolumn{2}{c}{Mg~{\sc xii} Lyman~$\alpha$} &
\multicolumn{2}{c}{O~{\sc viii} Lyman~$\alpha$} \\
$\Phi / b$ & $\theta_{0}$ & EW & FWHM & EW & FWHM & EW & FWHM& EW & FWHM \\
(M$_{\odot}$ yr$^{-1}$) & (\textdegree) & (eV) & (km~s$^{-1})$ 
& (eV) & (km~s$^{-1}$) & (eV) & (km~s$^{-1}$) & (eV) & (km~s$^{-1}$) \\ \hline
1.0 & 60 & 160 & 25,000$^a$ & --$^b$ & --$^b$ & --$^b$ & --$^b$ & --$^b$ & --$^b$ \\
1.0 & 30 & 130 & 22,000$^a$ & --$^b$ & --$^b$ & --$^b$ & --$^b$ & --$^b$ & --$^b$\\
3.0 & 60 & 310 & 17,000$^a$ & 8 &  4,800  & $\leq 3$ & --$^b$ & --$^b$ &--$^b$ \\
3.0 & 30 & 320 & 19,000$^a$ & 8 & 4,000  & 3 & 4,300 & 0.5 & 2,300 \\
6.0 & 60 & 370 & 18,000$^a$ & 18 & 3,800 & 7 & 3,300 & $\leq 2$ &--$^b$ \\
6.0 & 30 & 350 & 18,000$^a$ & 17 & 2,800 & 7 & 2,800 & 2 & 1,500\\
10.0 & 60 & 320 & 17,000$^a$ & 24 &  3,300  & 8 & 3,300 & 3 & 1,500 \\
10.0 & 30 & 340 & 18,000$^a$ & 23 &  2,500  & 9 & 2,000 & 4 & 1,800 \\
\multicolumn{2}{c}{Observations} & $95 \pm 20$ & $\sim 12,000$ & $32 \pm 12$ & &
$15 \pm 6$ & & $4.2 \pm 0.9$ & $< 2,000$ \\ \hline 
\end{tabular}

\noindent $^a$ To obtain a single FWHM for the Fe blend at 7.5~keV, the computed spectrum was
convolved with a Gaussian of FWHM = 10,000~km~s$^{-1}$. The FWHM reported in the table is that
measured from the convolved spectrum.

\noindent $^b$ In some cases, several of the lines do not appear as sharp spectral 
features at the terminal velocity
of the flow but are either very weak, broad features or absent. In such cases no values
are given in the table.

\end{table*}

The absolute value of the ionization fractions at the outer boundary of the 
wind is primarily determined by the density, a
consequence of the requirement placed on the models that the outcoming X-ray 
flux is consistent with the observed flux. Higher densities favour lower
ionization stages, as one would expect.

The model with $\Phi / b = 1.0 $~M$_{\odot}$ yr$^{-1}$ and 
$\theta_{0} = 90$\textdegree~shows very little variation of ionization with 
$r$ (top left panel of Figure~7). When the density is increased (top
right panel of Figure~7) the radiation energy density, and therefore 
photoionization rate, vary more significantly within the model. This occurs
because the higher density leads to a larger opacity (most importantly
due to electron scattering) meaning that photons take longer to escape the 
inner regions. The resulting gradient of photoionization rate with $r$ 
leads to a gradient of ionization with the highest ionization state in the 
inner wind.

Similarly, when the opening angle is reduced (lower left panel of Figure~7)
a gradient of ionization is introduced. This occurs because of photon leakage
through the conical boundary of the flow: many photons which contribute to 
the energy density in the inner flow leave the domain of the calculation
via the conical boundary before reaching the outer parts and therefore never
contribute to the energy density at large radii. This physical 
effect is solely due
to the geometry and is an important discriminant between
spherical and conical outflow in this, and potentially other, astrophysical
applications.

When a narrow opening angle
and high density are combined a very dramatic gradient of ionization is 
produced (lower right panel of Figure~7).

These gradients of ionization are responsible for the narrower line profiles
at high densities and small opening angles: a steep
gradient concentrates the line opacity in the outer part of the wind making
the lines sharper. Differences in the relative strengths of the Fe~{\sc xxvi}
and {\sc xxv} features at a given $\Phi/b$ are also the result of the 
sensitivity of the ionization gradient to $\theta_{0}$ since steep
gradients favour Fe~{\sc xxv} over Fe~{\sc xxvi} in a smaller volume
than shallow gradients.

Photon leakage from the flow is also responsible for changes in the continuum shape
at high energies. Comparing the dotted lines (electron scattering only) in Figures 3, 4 and 5,
it can be seen that as the opening angle is reduced the continuum curves less significantly
at high energies (this is particularly apparent for the highest density cases). 
The curvature of the continuum arises since Compton scattering preferentially removes high
energy photons and replaces them with low energy photons.
In spherical models, all the energy packets eventually emerge to contribute to the observed
spectrum therefore the maximum enhancement of the low energy intensity is seen for this
geometry. In conical models, however,  whenever a photon is scattered there is a chance that it will 
be directed out of the flow.
This leads to less enhancement of the low energy intensity and less curvature in the
continuum as seen by the observer. 

The equivalent widths and FWHM
of the important iron, sulphur, magnesium and oxygen lines computed from the bi-conical 
models are given in Table~3.
In the next section, these values will be compared with the observational constraints
reported by Pounds et al. (2003a).

\subsubsection{Confrontation with observations}

Compared with the spherical calculations presented in Section~5.1,
the spectra obtained from the bi-conical models are 
in better agreement with the
observations (Pounds et al. 2003a) primarily because, as discussed above,
reducing $\theta_{0}$ tends to produce deeper and narrower spectral lines.

As with the spherical models, it remains the case that the hard X-ray iron
feature (7.5~keV) is rather too strong in the model spectra 
(by about a factor of four
in equivalent width for most cases). It is possible that this discrepancy 
is, in part, due to the influence of the strong broad Fe K emission line
which is not modelled here but is believed to form by 
reflection from the inner accretion disk (Pounds et al. 2003a discuss
suitable fits for this feature).
The Fe~K emission line is immediately to the red of the observed absorption line
which leads
Pounds et al. (2003a) to remark on the 
possibility that
the absorption feature is only the Fe~{\sc xxvi} line and that the Fe~{\sc xxv}
component overlaps the Fe K emission feature. If this were the case then it 
would be expected that the observed equivalent width would be smaller than that
computed here. This may be particularly plausible at relatively high densities and
narrow opening angles since in such cases the two K~$\alpha$ absorption 
lines are individually rather narrow
and quite well separated in the model spectra.
It is also noted that any significant departure from solar abundances would directly influence the 
equivalent width.

The absorption at high energies ($\sim$9~keV) is less in the narrow opening angle models
than the spherical models. 
This is due to a combination of narrower line profiles and less continuum softening by Compton scattering
(see Section 5.2.2).
The high density models
($\Phi / b= 6$ and 10~M$_{\odot}$ yr$^{-1}$)
with 
$\theta_{0} = 60$\textdegree~still predict too much absorption in this region for 
consistency with the observations, but at $\theta_{0} = 30$\textdegree~the agreement
is closer.

Given the data quality and the relatively subtle differences between 
several of the model spectra it is not currently possible to
make a firm statement of a preferred model. However,
taking all the available observational constraints on PG1211+143 together, 
the spectra computed here suggest a moderately high density
in the flow (perhaps $\Phi / b \sim 6$~M$_{\odot}$ yr$^{-1}$) and a fairly narrow
opening angle ($\theta_0 \leq 30$\textdegree). High densities
are needed in order to produce the important soft X-ray features (to within
a factor of a few of their observed strengths) and a small opening angle
ensures narrow lines and limits the absorption at hard ($\sim 9$~keV) energy. 
Based on their detection of moderately strong
O~{\sc vii} emission in the RGS data, Pounds et al. (2003a) favoured a
relatively wide opening angle flow ($\sim 80$\textdegree). While the computations 
presented here do not rule out such a scenario they do suggest that
it will encounter difficulty in reproducing the observed narrowness of the 
X-ray features. 
Unfortunately, the models presented here do not predict any
O~{\sc vii} emission since the ionizing radiation field is too strong to permit
a significant population of this ion to exist in the flow. If the observed
O~{\sc vii} emission truly originates in the same material as the higher 
ionization state absorption lines, this would suggest an inaccuracy
of the ionization balance computed here. 
Alternatively, it is conceivable
that the emission arises, at least in part, from lower ionization material that is
physically distinct from the high
ionization outflow. A more complete discussion of 
O~{\sc vii} emission (and absorption) requires detailed investigation of the 
ionization/recombination balance in oxygen which is deferred to later work.

\section{Conclusions}

It has been shown that realistic radiative transfer calculations
support the proposal that the recently identified
X-ray absorption features in PG1211+143
(in particular the strong absorption line observed
at $\sim 7$~keV) can be explained
in terms of a simply parameterised bi-conical outflow model. 

Spectra have been computed for outflows with a range of both 
mass-loss rate and opening angle. 
The opening angle determines the extent of photon leakage from the flow which plays 
an important role in establishing a gradient of ionization in the wind.
This affects the strengths of the spectral features providing a useful discriminant 
between spherical and conical outflow.
The calculations presented
here favour models in which the flow has a moderately small opening angle 
$\leq 30$\textdegree~($b \leq 0.13$) and mass-loss rate 
a little larger than 
that suggested by Pounds et al. (2003a)~($\Phi/b \sim $ 6~M$_{\odot}$ yr$^{-1}$
is proposed here).

A narrow opening angle suggests interpretation in terms of a collimated disk wind or
jet originating from the accretion disk. Such outflows are certainly favoured as 
elementary components of the structure of quasars and a relatively narrow opening angle
bi-conical outflow is consistent with existing models for quasars (e.g.
Antonucci \& Miller 1985; Elvis 2000, 2004).
In particular, in the context of the Elvis (2000) model,
one may speculate that the highly ionized outflow modelled here lies in 
some portion of the polar region bounded by the WHIM.
Alternatively 
the X-ray features may
form in the most extremely ionized parts of the WHIM itself.

Given that high velocity outflows may be important for understanding the
processes which operate in the cores of quasars, there is a real need for 
higher quality observational data (such as may be provided by the next 
generation of X-ray observatories as discussed by Chartas et al. 2003).
In particular, it is to be hoped that an increase in spectral resolution
such as will be provided by the forthcoming {\it Astro-E2} mission,
will allow the study of line profiles in much greater detail than is currently
feasible.

As better observational data becomes available,
future theoretical work will include a detailed investigation of the various absorption
and emission features observed at softer energies. Although the O~{\sc viii}
Lyman~$\alpha$ line has been discussed here, several of the lower ionization
features that are observed (e.g. those of O~{\sc vii}) are not predicted
by the models at present. An important step in modelling these 
features is likely to be self-consistent 
calculation of the temperature in the models since variation of the 
temperature through 
the flow may lead to quantitative changes in the ionization balance owing 
to its effect on recombination rates. 
Temperature calculations will also provide insight into the formation of a
photosphere in the outflow, which might explain the BBB (Pounds et al. 2003a, King \& Pounds),
and help constrain the launching radius of the wind ($R_c$).
Such calculations go beyond the scope of
the simple modelling presented here but the influence of the choice of
temperature on the hard X-ray spectrum is briefly discussed in the Appendix.
It may also become necessary to consider angular ($\theta$) variation of the
ionization fractions in the flow, departures from ionization equilibrium,
rotation and the influence of viewing angle
if the basic bi-conical geometry stands up to
tighter observational constraints.

This paper has not addressed the important issue of how high velocity
flows may be accelerated. It has been suggested that radiation pressure
may be primarily responsible (King \& Pounds 2003, Pounds et al. 2003a).
Since the source here has been assumed to radiate at the Eddington luminosity
it is, by definition, the case that radiation pressure is sufficient to 
overcome gravity but details of how the flow is accelerated 
over and above gravity to the terminal velocity warrant investigation. 
Everett \& Ballantyne (2004) have concluded that continuum driving alone cannot account for 
outflows such as have been proposed by Pounds et al. (2003a). However, the
part played by spectral lines in driving an outflow has yet to be quantified. 
If the flow is radiatively driven by lines and bound-free continua then understanding
the driving is closely coupled to modelling the softer parts of the spectrum than
have been considered here: the X-ray region is relatively sparse of spectral lines and
very preliminary indications from the models presented here suggest that the lines in this
region are unlikely to provide a significant fraction of the required outward force.

In conclusion, the spectral synthesis presented in this paper suggests that
simple flow models can reproduce the absorption features observed in the hard
X-ray spectrum of PG1211+143. However, significant further study, 
both observational and theoretical, is needed in order that these high velocity
outflows and their relationship to the black hole accretion 
process which lies at the heart of the quasar phenomenon can be understood. 

\section*{Acknowledgements}

I thank L. Lucy for many useful discussions and suggestions regarding all aspects of this
work and the treatment of Compton scattering in particular. Thanks also to J. Drew and
K. Nandra from their advice and encouragement; to A. L. Longinotti for helpful discussions
regarding the interpretation of {\it XMM-Newton} data; and to an anonymous referee for
useful comments.
This work was carried out while I was a PPARC-supported PDRA at Imperial College London
(PPA/G/S/2000/00032).

\section*{References}
Antonucci R. R. J., Miller J. S., 1985, ApJ, 297, 621\\
Bautista M. A., Kallman T. R., 2001, ApJS, 134, 139\\
Boroson T. A., 2002, ApJ, 565, 78\\
Chartas G., Brandt W. N., Gallagher S. C., Garmire G. P.,\\
\indent ApJ, 2002, 579, 169\\
Chartas G., Brandt W. N., Gallagher S. C., 2003, ApJ, 595, \\
\indent 85\\
Elvis M., 2000, ApJ, 545, 63\\
Elvis M., 2004, in ``AGN Physics with the Sloan Digital \\
\indent Sky Survey'', ASP Conference Series 311,  eds. \\
\indent Richards G. T., Hall P. B., San Francisco: ASP, p.109\\
Everett J. E., Ballantyne D. R., 2004, to appear in ApJL,\\
\indent astro-ph/0409409\\
Gierli\'{n}ski M., Done C., 2004, MNRAS, 349, L7\\
Groenewegen M. A. T., Lamers H. J. G. L. M., 1989, \\
\indent A\&AS, 79, 359\\
Ignace R., Quigley M. F., Cassinelli J. P., 2003, ApJ, 596,\\
\indent 538\\
Kaspi S., Smith P. S., Netzer H., Maoz D., Jannuzi B. T., \\
\indent Giveon U., 2000, ApJ, 533, 631\\
Kaspi S., 2004, to appear in ``The Interplay among Black\\
\indent Holes, Stars and ISM in Galactic Nuclei'', IAU\\
\indent Symposium 222, eds. Bergmann Th. S., Ho L. C. \& \\
\indent Schmitt H. R., astro-ph/0405563\\
King A. R., Pounds K. A., 2003, MNRAS, 345, 657\\
Leighton R. B., 1959, ``Principles of Modern Physics'', (New \\
\indent York: McGraw-Hill), p. 433\\
Lucy L. B., 1999, A\&A, 345, 211\\
Lucy L. B., 2002, A\&A, 384, 725\\
Lucy L. B., 2003, A\&A, 403, 261\\
Marziani P., Sulentic J. W., Dultzin-Hacyan D., Calvani\\
\indent M., Moles M., 1996, ApJS, 104, 37\\
McKernan B., Yaqoob T., Reynolds C. S., 2004, to appear\\
\indent in ApJ, astro-ph/0408506\\
Mitsuda K. et al., 1984, PASJ, 36, 741\\
Pier E. A., Krolik J. H., 1992, ApJ, 399, L23\\
Pounds K. A., Reeves J. N., King A. R., Page K. L., \\
\indent O'Brien P. T., Turner M. J. L., 2003a, MNRAS, 345, \\
\indent 705\\
Pounds K. A., King A. R., Page K. L., O'Brien P. T., \\
\indent MNRAS, 2003b, 346, 1025\\ 
Reeves J. N., O'Brien P. T., Ward M. J., 2003, ApJ, 593, \\
\indent L65\\
Sim S. A., 2004, MNRAS, 349, 899\\ 

\section*{Appendix: The influence of electron temperature}

Throughout the calculations presented in this paper an electron temperature
$T_{e} = 1.3 \times 10^6$~K has been adopted. This temperature was chosen
based on the primary black-body temperature obtained from fits
to the X-ray spectrum of PG1211+143 by Pounds et al. (2003a) and is
close to the temperature used in that paper for the analysis of the
observed O~{\sc vii} emission feature.

In principle, the electron temperature in the outflow can be determined by
detailed consideration of heating and cooling mechanisms but such 
computations go beyond the scope this paper which is primarily concerned
only with radiative transfer and spectral synthesis.

It is instructive, however, to consider how different choices of temperature
may influence the computed spectrum. To this end, Figure~8 shows spectra
computed for spherical models with $\Phi = 3.0$~M$_{\odot}$ yr$^{-1}$ and 
temperatures both above 
($T_{e} = 2.6 \times 10^{6}$~K) and below ($T_{e} = 6.5 \times 10^{5}$~K)
the standard temperature. Computations at the standard temperature are
also shown, for comparison.
Values of the equivalent widths and FWHM of the 7.5~keV iron feature
and the Lyman $\alpha$ line of hydrogen-like sulphur in these
spectra are given in Table~4.

\begin{figure*}
\vspace*{-4cm}
\epsfig{file=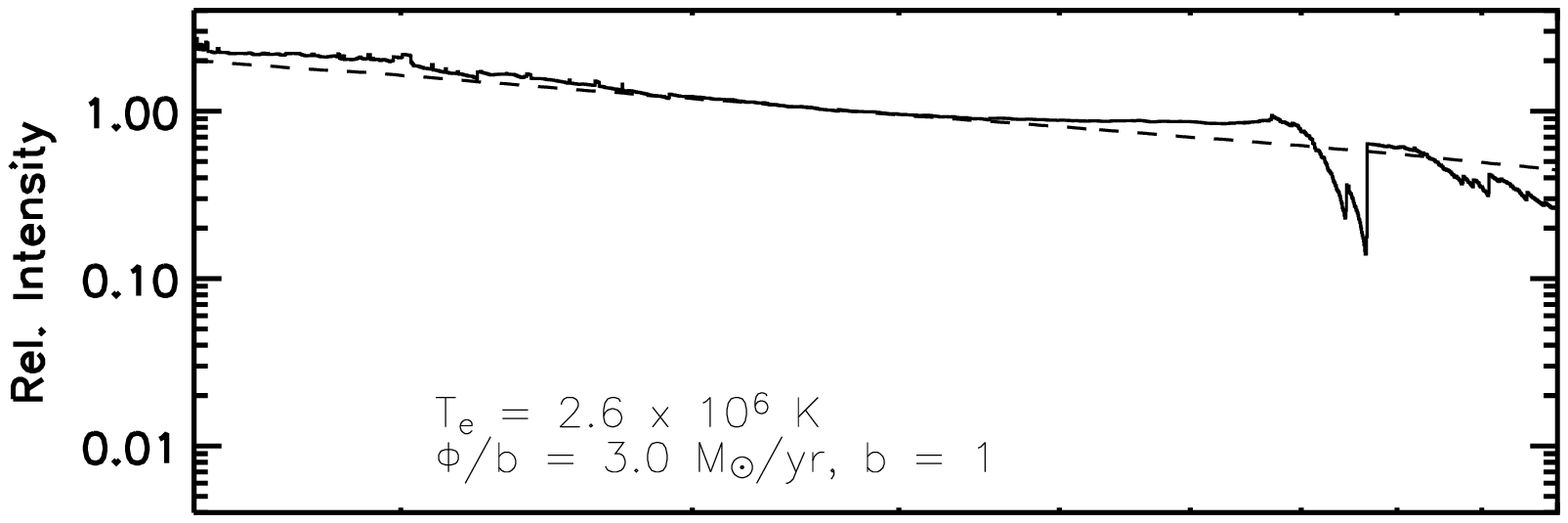, width=12cm, height=8cm}\\
\vspace*{-4cm}
\epsfig{file=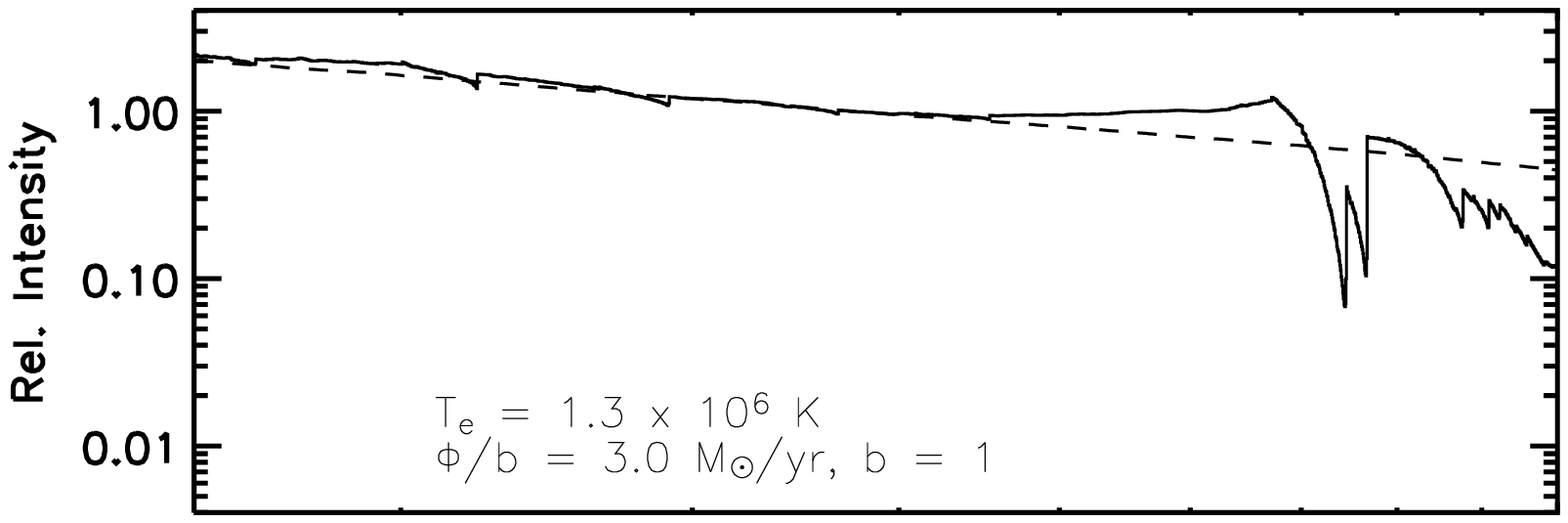, width=12cm, height=8cm}\\
\vspace*{-4cm}
\epsfig{file=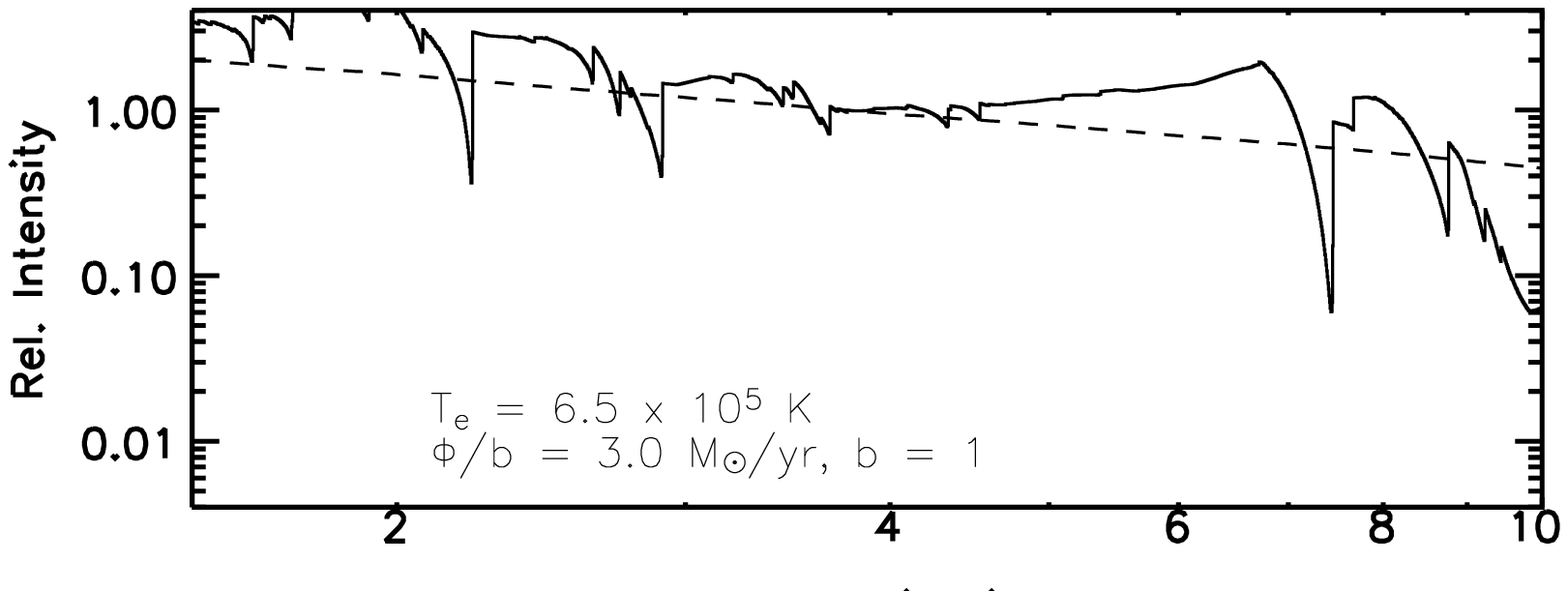, width=12cm, height=8cm}\\
\caption{2 -- 10 keV spectra computed from spherical models with 
$\Phi = 3$~M$_{\odot}$ yr$^{-1}$ 
for different electron temperatures ($T_{e}$).
The panels show spectra computed with $T_{e} = 2.6 \times 10^6$~K (upper panel),
$T_{e} = 1.3 \times 10^6$~K (middle panel) and $T_{e} = 6.5 \times 10^5$~K
(lower panel). 
The dashed lines show continuum spectra, computed in the absence of
bound-free processes and spectral lines.}
\end{figure*}

\begin{table*}
\caption{Equivalent width (EW) and full-width at half-maximum (FWHM) for the 
computed absorption components
of the 
S~{\sc xvi} Lyman~$\alpha$ line and the combined Fe~{\sc xxvi}/{\sc xxv}
feature at 7.5~keV for spherical wind models 
($\theta_{0} = 90$\textdegree, $b=1$) with 
$\Phi = 3$~M$_{\odot}$ yr$^{-1}$ and different temperatures. 
The Monte Carlo simulations are accurate to about $\pm 10$ per cent in EW 
and $\pm 300$~km~s$^{-1}$ in FWHM. 
}
\begin{tabular}{lcccc} \hline
{Model} & \multicolumn{2}{c}{Fe (7.5 keV)} & 
\multicolumn{2}{c}{S~{\sc xvi} Lyman~$\alpha$} \\
$T_{e}$ & EW & FWHM & EW & FWHM\\
(K) & (eV) & (km~s$^{-1})$ 
& (eV) & (km~s$^{-1}$) \\ \hline
$2.6 \times 10^6$ & 250 & 18,000$^a$ & --$^b$ & --$^b$ \\
$1.3 \times 10^6$ & 360 & 21,000$^a$ & 10& 7,400 \\
$6.5 \times 10^5$ & 370 & 21,000$^a$ & 21 & 7,500 \\ \hline
\end{tabular}

\noindent $^a$ To obtain a single FWHM for the Fe blend at 7.5~keV, the computed spectrum was
convolved with a Gaussian of FWHM = 10,000~km~s$^{-1}$. The FWHM reported in the table is that
measured from the convolved spectrum.

\noindent $^b$ In this case, the sulphur line is 
too weak for reliable calculation above the Monte Carlo noise.

\end{table*}

At higher temperatures the spectral features become weaker while at
lower temperatures they become stronger. This is due to the 
influence of the temperature on the recombination rates -- at high
temperatures the flow is more highly ionized than at low temperatures.
It suggests a degeneracy between high densities and 
low temperatures but this is lifted when the linewidths are considered --
high densities tend to lead to narrower lines (see Table~2) because
of the ionization gradient they introduce (see Section~5.2.2) but a
low temperature does not change the gradient of ionization and therefore
does not cause the lines to become narrower. 
This means that a simple change in the temperature cannot simultaneously
address the shortcomings of the low density spherical models discussed 
in Section 5.1 (i.e. changing the adopted value of $T_{e}$ will not cause 
the lines to be both narrower and stronger). However, the temperature
does have a quantitative effect on the lines via its influence on the
ionization balance.

In the future, calculations will be presented in which the temperature
is computed by balancing radiative heating and cooling processes, following
e.g. Lucy (2003). Such work will allow for the interesting possibility of
variations of $T_{e}$ with $r$: if present, this would
affect the gradient of ionization which is central to obtaining narrow line
profiles in the models discussed here. It is possible that a steep
gradient of electron temperature would permit lower ionization stages to form
in the outermost parts of the flow, thereby helping to explain many of 
the lines observed in the RGS spectrum of PG1211+143.

\label{lastpage}
\end{document}